\documentclass[useAMS,usenatbib]{mn2e}
\pdfoutput=1
\usepackage[outercaption]{sidecap}    
\usepackage{graphicx}
\usepackage{color}
\definecolor{highlight}{rgb}{1,0,0}
\definecolor{note}{rgb}{0,0,1}
\usepackage{caption}
\usepackage{paralist} 
\usepackage{pdfpages}
\usepackage{aas_macros}
\usepackage{amssymb}
\usepackage{pdfpages}
\usepackage{lscape}
\usepackage{rotating}

\def\eagle{{\sc eagle}}
\newcommand{\x}{$\chi^2$}
\newcommand{\msr}{$\mathcal{M_\star}-R_{\rm e}$}
\newcommand{\bt}{$\rm B/T$}
\newcommand{\galfit}{\textsc{galfit}}
\newcommand{\bd}{bulge + disc$\ $}

\DeclareGraphicsExtensions{.pdf,.png}
\graphicspath{{plots/}}
\DeclareCaptionFormat{cont}{#1 (cont.)#2#3\par}

\title[GAMA: $z=0$ component \msr{} relations]
	{Galaxy And Mass Assembly (GAMA): \msr{} relations of $z=0$ bulges, discs and spheroids}
\author[R. Lange et al.]
{Rebecca Lange$^{1}$\thanks{email:rebecca.lange@icrar.org}, Amanda J. Moffett$^1$, Simon P. Driver$^{1,2}$, Aaron S.G. Robotham$^{1}$,  
\newauthor  Claudia del P. Lagos$^{1,3}$, Lee S. Kelvin$^{4}$, Christopher Conselice $^5$, 
\newauthor  Berta Margalef-Bentabol$^5$, Mehmet Alpaslan $^6$, Ivan Baldry$^{4}$, Joss Bland-Hawthorn$^7$, 
\newauthor   Malcolm Bremer$^8$, Sarah Brough$^9$, Michelle Cluver$^{10}$, Matthew Colless$^{11}$, 
\newauthor Luke J. M. Davies$^1$,  Boris H{\"a}u{\ss}ler$^{12}$, Benne W. Holwerda$^{13}$, Andrew M. Hopkins$^7$, 
\newauthor Prajwal R. Kafle$^1$, Rebecca Kennedy$^5$, Jochen Liske$^{14}$, Steven Phillipps$^{8}$, 
\newauthor Cristina C. Popescu$^{15,16}$, Edward N. Taylor$^{17}$, Richard Tuffs$^{18}$, Eelco van Kampen$^{19}$, 
\newauthor Angus H. Wright$^1$ \\	      
$^1$International Centre for Radio Astronomy Research, University of Western Australia, M468, 35 Stirling Highway, Crawley, WA6009, Australia\\
$^2$Scottish Universities' Physics Alliance (SUPA), School of Physics and Astronomy, University of St. Andrews,\\ North Haugh, St. Andrews, KY16 9SS, UK\\
$^3$Australian Research Council Centre of Excellence for All-sky Astrophysics (CAASTRO), 44 Rosehill Street, Redfern, NSW2016, Australia\\
$^4$Astrophysics Research Institute, Liverpool John Moores University, IC2, Liverpool Science Park, 146 Brownlow Hill, Liverpool, L3 5RF, UK \\
$^5$ School of Physics and Astronomy, The University of Nottingham, University Park, Nottingham, NG7 2RD, UK\\
$^6$NASA Ames Research Center, N232, Moffett Field, Mountain View, CA 94035, United States\\
$^7$Sydney Institute for Astronomy, School of Physics A28, University of Sydney, NSW 2006, Australia\\
$^{8}$Astrophysics Group, School of Physics, University of Bristol, Bristol BS8 1TL, UK\\
$^9$Australian Astronomical Observatory, PO Box 915, North Ryde, NSW 1670, Australia\\
$^{10}$Department of Physics and Astronomy, University of the Western Cape, Robert Sobukwe Road, Bellville, 7535, South Africa\\
$^{11}$Research School of Astronomy and Astrophysics, Australian National University, Canberra, ACT 2611, Australia\\
$^{12}$European Southern Observatory, Alonso de Cordova 3107, Vitacura, Santiago, Chile\\
$^{13}$University of Leiden, Sterrenwacht Leiden, Niels Bohrweg 2, NL-2333 CA Leiden, The Netherlands \\
$^{14}$Universit\"at Hamburg, Hamburger Sternwarte, Gojenbergsweg 112, D-21029 Hamburg, Germany \\
$^{15}$Jeremiah Horrocks Institute, University of Central Lancashire, Leighton Building, Preston PR1 2HE, United Kingdom \\
$^{16}$The Astronomical Institute of the Romanian Academy, Str. Cutitul de Argint 5, Bucharest, Romania\\
$^{17}$School of Physics, the University of Melbourne, VIC 3010, Australia\\
$^{18}$Max-Planck-Institut f\"ur Kernphysik, Saupfercheckweg 1, D-69117 Heidelberg, Germany\\
$^{19}$European Southern Observatory, Karl-Schwarzschild-Str.2, D-85748 Garching, Germany
}

\pagerange{\pageref{firstpage}--\pageref{lastpage}}\pubyear{2016}
\begin{document}
\label{firstpage}
\maketitle
\begin{abstract}
We perform automated \bd decomposition on a sample of $\sim$7500 galaxies from the 
Galaxy And Mass Assembly (GAMA) survey in the redshift range of 0.002$<$z$<$0.06 
using \textsc{sigma}, a wrapper around \galfit3. 
To achieve robust profile measurements we use a novel approach of repeatedly
fitting the galaxies, varying the input parameters to sample a large fraction 
of the input parameter space.
Using this method we reduce the catastrophic failure rate significantly and verify the confidence in the fit independently of \x. Additionally, using the median of the final fitting values and the 16$^{th}$ and 84$^{th}$ percentile produces more realistic error estimates than those provided by \galfit, which are known to be underestimated.\\
We use the results of our decompositions to analyse the stellar mass -- half-light radius 
relations of bulges, discs and spheroids. We further investigate the association of components 
with a parent disc or elliptical relation to provide definite $z=0$ disc and spheroid 
\msr{} relations.
We conclude by comparing our local disc and spheroid \msr{} to simulated data
from \eagle\ and high redshift data from CANDELS-UDS. We show the potential of using the
mass-size relation to study galaxy evolution in both cases but caution that for a fair 
comparison all data sets need to be processed and analysed in the same manner. 
\end{abstract}
\begin{keywords}
galaxies: fundamental parameters - galaxies: statistics - galaxies: formation - galaxies: elliptical and lenticular - galaxies: spiral
\end{keywords}

\section{Introduction}

At the fundamental level galaxies are multi-component systems \citep[see for example][]{Buta2010},
consisting of at least a spheroid and/or disc. This is most obvious in the 
S\'ersic index -- colour plane where the single component Sd 
and elliptical galaxies occupy distinct peaks with composite galaxies (S0abc) 
scattered between and around these peaks \citep[see e.g.,][]{Driver2006,Cameron2009,Kelvin2012,Lange2015}.
These components have very different characteristics with spheroids typically 
having a featureless appearance and being pressure supported.
Discs on the other hand have features such as spiral arms and are rotationally supported.
Furthermore bulges are made up of redder stars with moderate to high 
metallicities and a high $\alpha$-element abundance, while discs are made of 
younger, bluer stars with lower metallicities and typically are dust and gas rich. 
Spheroids are older, showing little to no star formation and are 
typically dust and gas depleted \citep[see for example the review by][]{RobertsHaynes1994}. 
The simplest explanation for these stark differences is that spheroids and discs form via two distinct  
mechanisms over two distinct eras \citep{Cook2009,Driver2013}, i.e. a dynamically \mbox{``hot mode''} 
(spheroid formation) and ``cold mode'' (disc formation) evolution.\\

Traditionally the relative prominence of a bulge component is taken into account when 
classifying galaxies onto the Hubble sequence (see \citealt{Hubble1926}, 
and later revisions by e.g., \citealt{vdBergh1976,Kormendy2012}), however studying global 
properties of galaxies by Hubble type could be misleading. 
For example, numerous evolution mechanisms have been proposed to explain the morphological diversity 
seen at $z=0$, such as a (initial) major dissipative event, gas accretion, adiabatic contraction, 
major and minor mergers and secular processes 
\citep[see e.g.,][]{Hopkins2010,Trujillo2011,L'Huillier2012,Cheung2013,Sachdeva2015}.
Each of these processes potentially acts to modify the prominence of the bulge, disc
or other components. This indicates that galaxy components likely
follow distinct formation pathways and structure effectively encodes the formation history.
Therefore, to study galaxy evolution \bd{} decomposition is critical.

While the number of studies of large samples which employ \bd{} 
decomposition to explore the nature of galaxies and their components is growing, the analysis is challenging 
\citep[see e.g.,][]{Allen2006,Gadotti2009,Simard2011,Bruce2012,Bruce2014,Lang2014,Tasca2014,Meert2015,Salo2015}. 
This is because multi-component fitting is notoriously difficult, especially when trying to
automate it for large samples. Nevertheless a number of publicly available codes have now 
been created to allow \bd{} decomposition, such as
\textsc{gim2d} \citep{Simard1998}, \textsc{budda} \citep{deSouza2004}, \textsc{galfit3} 
\citep{Peng2010} and \textsc{imfit} \citep{Erwin2015}.
Each code has advantages and disadvantages \citep[see][for example for further discussion]{Erwin2015}, 
here we elect to use \galfit 3 because of its ability to manage nearby objects, 
its computational reliability, and its speed.

Many studies that fit 2-component S\'ersic light profiles restrict 
the S\'ersic index to n=1 for the disc and in some cases n=4 for the bulge
\citep[e.g.~][]{Simard2011,Bruce2012,Lackner2012,Meert2015}. 
This reduces the number of free parameters and ensures the fitting process is more 
robust but it restricts the possible interpretations of the fitting outcomes, 
e.g. classical and pseudo bulges can not be differentiated this way. A number of
studies now show that the S\'ersic index of discs and spheroids (be they pure or component)
vary smoothly with mass and luminosity or due to dust or galaxy type 
\citep[see e.g.][]{Graham2003,Gadotti2009,Kelvin2012,Graham2013,Pastrav2013a,Pastrav2013b}. 
Hence studies where the S\'ersic index of the bulge or disc components are 
fixed may be overly restrictive.  
Furthermore, to correctly trace a galaxy's formation history a full decomposition of all of its components 
would be ideal \citep[e.g., the Spitzer Survey of Stellar Structure in Galaxies, S$^4$G,][]{Salo2015}. 
However, this is only viable for very nearby galaxies where all the components can be clearly 
resolved and hence for relatively small samples (S$^4$G is the largest study to date 
extending to 2352 galaxies for which a number have been fit with more than 2 components).
To compare to galaxies at different epochs going beyond a simple bulge and disc 
decomposition is difficult \citep{Gadotti2008}. 
There are two reasons, however, why two components might be sufficient, (i) the 
majority of stellar mass resides in the bulge and disc components for most galaxies, and
(ii) some components may simply represent minor perturbations to the underlying disc (e.g. 
bars, pseudo-bulges). Such perturbations should arguably be considered secondary rather 
than primary evolutionary markers. 

Here we adopt the stance that bulge and disc components arise from two primary formation 
pathways (i.e. hot and cold mode evolution, respectively), and that additional 
components form in secondary formation pathways (i.e., tidal interactions, 
disc instabilities and perturbations).
The likely primary pathways are: monolithic collapse followed by major mergers, 
which can produce elliptical galaxies by 
destroying and rearranging any structure previously present in a galaxy, resulting in a 
smooth light profile \citep{Toomre1977}; and minor mergers and continued gas inflow, 
which can form or re-grow a disc around a pre-existing spheroid, resulting in a galaxy 
with two distinct components \citep[see e.g.,][]{Steinmetz2002,Kannappan2009,Wei2010}. 
A key question worth asking is whether two generic components
(spheroids and discs) really can explain the diversity seen, i.e., 
how many fundamental building blocks and structures are required to adequately reproduce the observed galaxy population?  As most of the stellar mass is contained within the bulge and disc 
how important are tertiary features like bars? Furthermore how many different physical origins 
do the various spheroids and discs have?
Are elliptical galaxies simply naked bulges and are bulges related to high-redshift compact galaxies \citep[e.g.][]{Graham2015,Berg2014}? 
Are the discs of early-types, late-types and irregulars indistinguishable? \\

We believe that the stellar mass -- half-light size (hereafter \msr{}) relation is a key scaling 
relation allowing us to address these questions for the following reasons:
\begin{itemize}
\item	The size of a galaxy is related to its specific angular momentum making the mass and size of a galaxy fundamental observables of conserved quantities \citep[e.g.,][]{Romanowsky2012}.
\item	The simple assumption that angular momentum is conserved during the initial collapse of the dark matter halo links the
angular momentum and mass of a galaxy with its dark matter halo \citep{Fall1980,Dalcanton1997,Mo1998}.
\item	Hydrodynamical simulations now produce galaxies with realistic sizes and direct comparisons (at different epochs) are possible to study formation and evolution histories of galaxies  \citep[see for example the Evolution and Assembly of Galaxies and their Environments simulation suite, \eagle,][]{Schaye2015,Crain2015}.
\item	We can empirically measure and trace the masses and sizes of galaxies and their components over a range of redshifts and in different environments (e.g., with HST as well as high-redshift ground-based 
surveys and soon with Euclid and WFIRST).

\end{itemize}

The \msr{} relation therefore represents the next critical diagnostic for galaxy evolution
studies beyond simple mass functions \citep[see e.g.,][]{Bouwens2004,Wel2014,Holwerda2015,Shibuya2015},
 enabling us to trace angular momentum build-up and the 
emergence of the component nature of galaxies while connecting observations to simulations.\\

Recent studies comparing the \msr{} relation of low and high-redshift are 
already yielding interesting results.
For example, at high-redshift galaxies might look disc-like or elliptical\slash spheroidal 
but their physical properties are unlike any discs or ellipticals in the local Universe 
\citep[see e.g.~][]{Bruce2012,Buitrago2013,Mortlock2013}.
Galaxies at high redshifts are typically more irregular with thick slab-like disc structures and
clumpy star-forming regions \citep{Wisnioski2012}. 
In addition, they can be very compact but massive. 
In some cases, at redshift $\sim$2 they are a factor of up to 6 times smaller in size than galaxies 
of the same mass today \citep{Daddi2005,Trujillo2007,Buitrago2008,vanDokkum2008,vanDokkum2010,Weinzirl2011}.\\

In this paper we aim to provide a reliable low redshift benchmark of the \msr{}
relation for bulges, discs and spheroids. 
The \bd decomposition sample
is derived from a set of galaxies for which detailed morphological
information is available (see \citealt{Moffett2015}).
Section \ref{sec:data} describes the data and sample selection, 
Sections \ref{sec:2comp} and \ref{sec:compmass} describe the set up of our 
\bd decomposition catalogue and component mass estimates. 
In Section \ref{sec:msr} we present the \msr{} relations for bulges, spheroidal 
and disc galaxies and discuss the association of components with their possible parent populations.
We then compare our distributions to the \eagle\ simulation in Section \ref{sec:sims} followed by a
comparison of our low redshift \msr{} relation with recent high redshift data from 
Ultra Deep Survey (UDS) region within the Cosmic Assembly Near-infrared Deep Extragalactic 
Legacy Survey \citep[CANDELS,][]{Grogin2011,Koekemoer2011} in Section \ref{sec:highz}.
Finally in Section \ref{sec:SnC} we present our summary and conclusions.\\

 Throughout this paper we use data derived from the Galaxy And Mass Assembly
(GAMA) survey \citep{Driver2011,Driver2016,Liske2014} with stellar
masses derived from \cite{Taylor2011}, sizes derived from S\'ersic profile fitting using \textsc{sigma}
\citep{Kelvin2012}, and for a cosmology given by: $\Lambda{}$ Cold Dark
Matter universe with $\Omega_{\mathrm{m}} = 0.3,\;
\Omega_{\Lambda}=0.7,\; H_0 = 70 \mathrm{km
  s}^{-1}\;\mathrm{Mpc}^{-1}$.

\section{Data}
\label{sec:data}

The GAMA survey is an optical spectroscopic and multi-wavelength
imaging survey of $\sim$ 300000 galaxies, combining the data of several ground and space based
telescopes \citep{Driver2011,Driver2016}. It is an
intermediate survey in respect to depth and survey area
\citep{Baldry2010} and thus fits in between low
redshift, wide-field surveys such as SDSS \citep{York2000} or 2dFGRS
\citep{Colless2003} and narrow deep field surveys like zCOSMOS
(\citealt{Lilly2007} and see \citealt{Davies2014}) or DEEP-2 \citep{Davis2003}. \\

In this paper we use data from the \mbox{GAMA II} \citep{Liske2014} equatorial regions, which
are centered on 9h (G09), 12h (G12) and 14.5h (G15).  The three
regions are 12$\times$5 $\mathrm{deg}^{2}$ in extent and have an \textit{r}-band
Petrosian magnitude limit of $r<$19.8 mag.  The spectroscopic target
selection is derived from SDSS DR 7 (\citealt{Abazajian2009}, see \citealt{Baldry2010} for details) 
input catalogue and we reach a spectroscopic completeness of $\ge$ 98\% for
the main survey targets. The redshifts
\citep{Baldry2014, Liske2014} are based on spectra taken with the AAOmega
spectrograph at the 3.9m Anglo-Australian-Telescope
\citep{Hopkins2013} located at Siding Spring Observatory.
The supporting panchromatic imaging data extends from the FUV to the far-IR
via GALEX, SDSS, VISTA, WISE and HerschelL \citep[for full details see][]{Driver2016}.
 All optical and near-IR imaging data has matched aperture photometry
\citep{Hill2011, Liske2014} 
Here we focus on the redshifts (\textit{SpecCatv27}), morphologies (\textit{VisMorphv03}), 
optical imaging (\textit{ApMatchedv06}) and stellar masses (\textit{StellarMassCatv18}) data products.

\subsection{Sample Selection}
\label{sec:selection}

\begin{figure}
\centering
\includegraphics[width=0.45\textwidth]{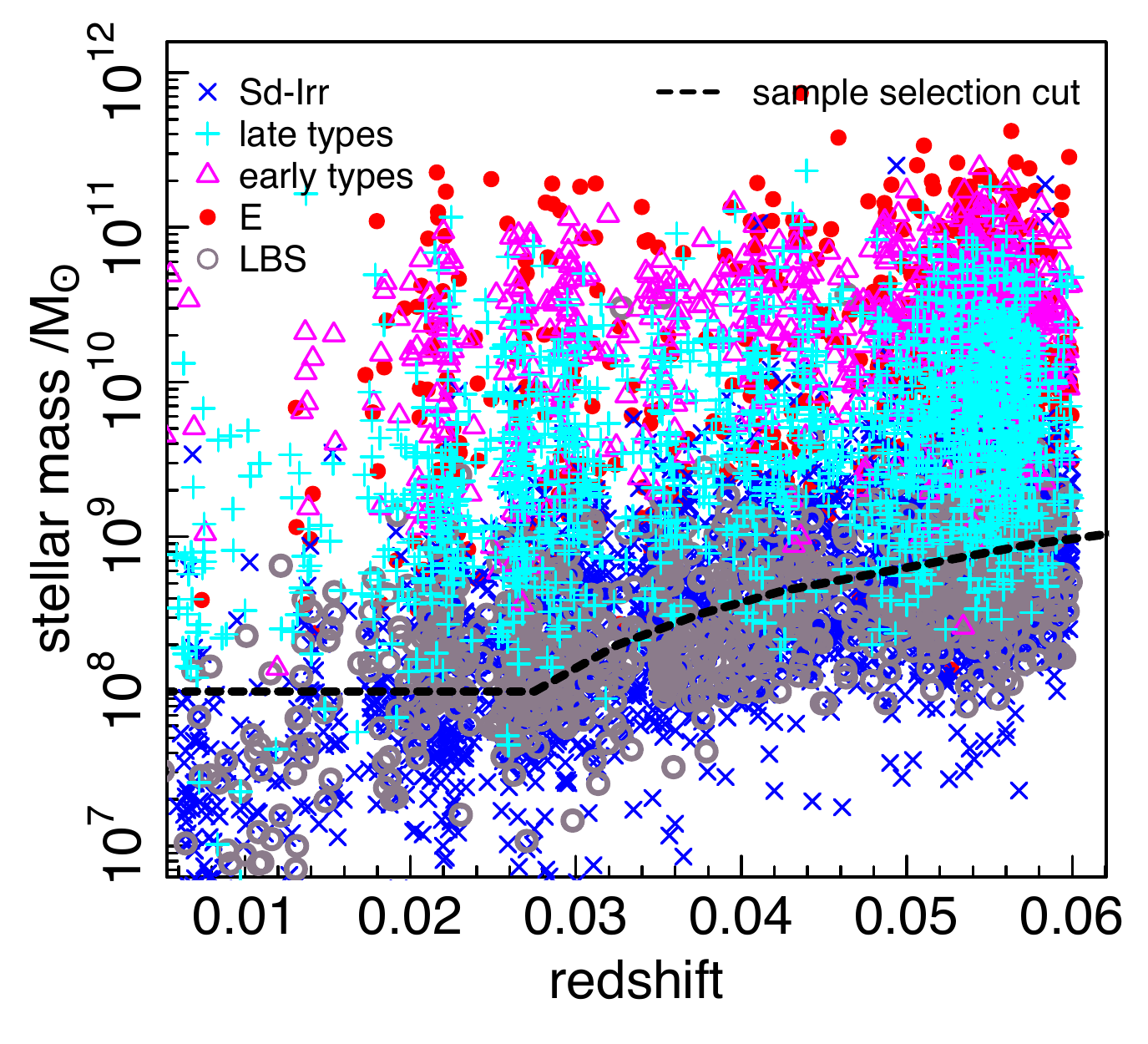}
\caption{Shown is the redshift -- stellar mass distribution for the GAMAnear sample. The points are coded according to the Hubble type established in the visual morphology classification. There appears to be no bias towards a particular Hubble type at the higher redshift boundary of our sample.}
\label{fig:morph_dist}
\end{figure}

We select galaxies with 0.002$<$z$<$0.06, \textit{r}$_{petro}<19.8$ mag and spectra quality NQ$>2$ (see \citealt{Liske2014}) for which visual morphologies have been established following \cite{Moffett2015}. 
To briefly summarise the visual classification procedure: we use
three-colour (\textit{Hig}) postage stamps of the objects to visually inspect them.
A simple classification tree is used to sort galaxies in the first
instance into bulge and disc dominated, little blue spheroid (LBS) and
star/artifact. In the following step the bulge and disc dominated objects 
are further split into single- and multi-component. Finally the
multi-component galaxies are sorted into barred and unbarred.  For
each galaxy the result was then translated to a Hubble type: E, S0-Sa,
SB0-SBa, Sab-Scd, SBab-SBcd and Sd-Irr, plus the additional LBS, Star
and Artifact classifications \citep[see][for further details]{Moffett2015}.
Note that throughout this paper the terms early- and late-type refer to our visual classification
of objects being bulge or disc dominated and we do not impose any other parameter cuts 
(e.g. S\'ersic index) to classify early- or late-type.\\

We perform \bd{} decomposition in the \textit{r}-band only as our fitting
approach is computationally expensive. For the analysis we do not consider 
objects classified as stars or artifacts and we excluded one additional galaxy
which was too large to be fit robustly.  
The resulting sample is hereafter called GAMAnear and comprises 7506
galaxies of which 2247 were visually classified as 2-component (S0-Sa,
SB0-SBa, Sab-Scd, SBab-SBcd) and 
5259 as single component (E, Sd-Irr, LBS) galaxies. Due to the low redshift range
this sample extends well below $10^9  \mathcal{M}_\odot$ allowing us to study the 
low-mass end of the \msr{} relation. However, because of the limited volume of our survey, 
this also means we do not have
many very high-mass galaxies to study the curvature of the \msr{} relation at higher masses.
This is important as the curvature in the (elliptical) \msr{} relation is likely indicative of the assembly history of the galaxy population (see e.g. \citealt{Bernardi2011} and references therein).\\

As shown by \citep[their Fig.~5]{Moffett2015} the fraction of galaxy type by mass behave as expected, e.g. the fraction of early-type galaxies increases with increasing mass and the fraction of late-type galaxies decreases. 
To check whether the sample selection in this paper is biased we 
 show the sample distribution in the redshift -- stellar mass
plane in Fig.~\ref{fig:morph_dist}.  The dashed line shows the final mass limit used 
in Sec.~\ref{sec:msr} to derive the \msr{} relation.
The Figure illustrates the mass segregation of the sample with the
early-type galaxies being more massive than the late-types. 
However, there is no clear bias with morphological type or our redshift range, 
especially at the upper redshift boundary. 

\newpage
\section{Bulge + Disc decomposition}
\label{sec:2comp}

Obtaining reliable \bd fits is notoriously difficult, particularly in
an automated fashion for large samples where the signal-to-noise ratio and
resolution varies. Typically 20 to 30$\%$ of automatic fits are in some way non-physical.
Previous studies have made use of a logical filter
(e.g., \citealt{Allen2006}, see also \citealt{Simard2011,Meert2015})  to identify
unphysical fits (e.g., component profiles which cross twice, the
switching of the bulge and disc components etc., see \citealt{Allen2006}
for more details) and manage these failures by replacement with a
single S\'ersic fit. As a first step this reduces the catastrophic
failure rate significantly but
introduces a bias by removing the subset of two-component systems with
poor fits. Following extensive exploration of our data using \galfit 3 \citep{Peng2010}, embedded in \textsc{sigma} \citep{Kelvin2012}, 
we identify five commonly occurring 
key factors which lead to poor and often catastrophic
fitting outcomes. These are summarised below along with our
adopted solution:\\

\noindent
{\bf (1) Becoming trapped in local minima and/or the limited movement
  of the final converged solution away from the initial conditions.}

\noindent
The Levenberg-Marquart (LM) \x{} minimisation algorithm used by \galfit 3 can
 get stuck in a local rather than the global \x{} minimum, especially when fitting multiple components.
One way to overcome this is to
vary the initial conditions (i.e., starting points) and repeat the fitting
process. Convergence to a common solution, regardless of the starting point, provides 
confidence that the true minimum has been found.
In due course a full Markov Chain Monte Carlo (MCMC) approach, that appropriately samples
the prior distribution should be developed but that is beyond the scope of our current
investigation at this stage.

~

\noindent 
{\bf (2) Unphysical solutions, e.g., a scale-length of 0.1$''$ or a
  S\'ersic index of 20.}

\noindent
In some cases the bulge or disc fits can migrate to the fit limits imposed,
 and these results are often not physical. 
 While it is tempting to reduce the limits to plausible values this causes
 a non-physical build-up of the solutions at the limits. Moreover during the path towards 
 convergence it can sometimes be seen that solutions migrate into extreme values and then back again.
To minimise the impact of our boundaries we imposed no limits on the parameters, bar the 
constraint on the centre position, which is set to $\pm$ 5 pixels to account 
for the oversampling of the PSF (i.e. GAMA pixel size is 0.339'' and 
SDSS FWHM=1.5''). Instead we elect to remove final solutions which settle on extreme values.
We can afford to do this since we have multiple fits for each galaxy, i.e.,
some starting points lead to extreme outcomes but on the whole most converge to plausible values.
Note that \galfit{} does have some inbuilt constraints, such as a maximum
S\'ersic index of 20.

~

\noindent
{\bf (3) Decision on single or multiple components.}

\noindent
A key problem in galaxy decomposition is to decide how many components are required.
Ideally this should be derivable from the independent 1-, 2- or multi-component fits.
Experimentation with the Akaike and Bayesian Information Criteria 
(AIC and BIC respectively) was explored but no obvious
automated process for determining the number of components, which
agreed with our visual assessments, was identified. This is in part due to
the limited information available in single band fitting.
Hence we adopt our visual classifications as priors, i.e., E, Sd-Irr and LBS
galaxies are taken as single component systems and S0-Sa,
SB0-SBa, Sab-Scd and SBab-SBcd as two-component. 
For completeness, however, we do derive and provide both 1 and 2 component fits for all systems.

~
 \begin{figure*}
\includegraphics[height=0.3\textheight]{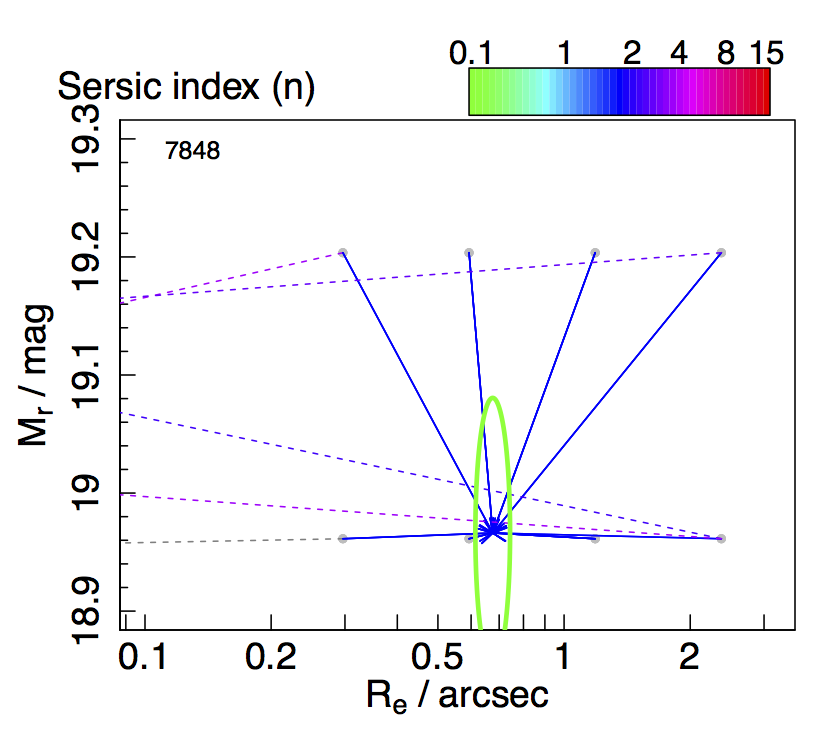}
\includegraphics[height=0.3\textheight]{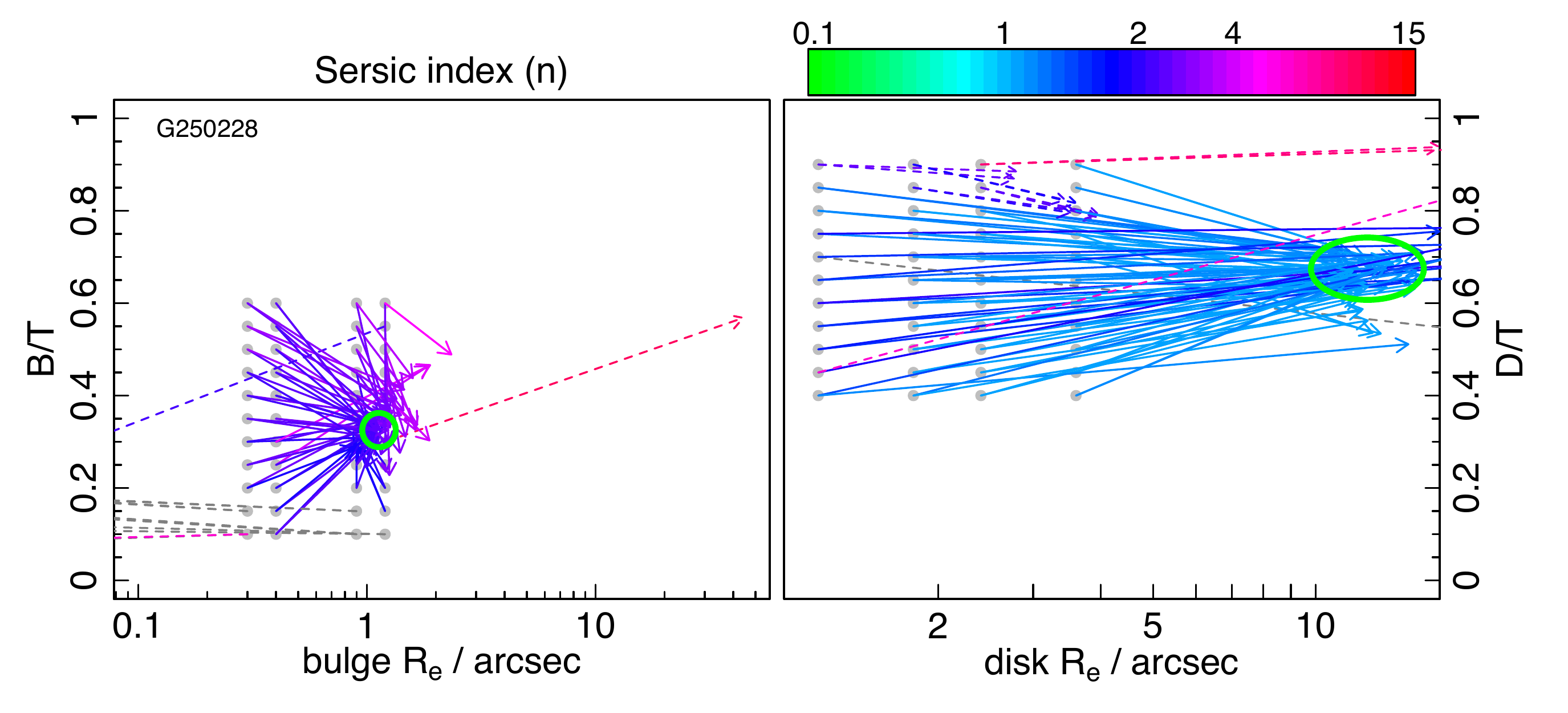}
\caption{Convergence plot examples for a single component galaxy in the top panel and a double component galaxy in the bottom panel. The plots show the starting (grey points) and end values (arrow head) for several fit parameters making it easy to evaluate how well the galaxy was fit. For the single component fits we show the galaxy's total magnitude, size and S\'ersic index. For the two-component fit we show the bulge-to-total (\bt) and disc-to-total ($\rm D/T$) flux ratios of the components instead of the magnitude. The green ellipse is centred at the median output values and its size corresponds to the adopted error on the median. Note that for the single component fit the error on the magnitude corresponds to our error floor of 0.11 mag. The dashed lines indicate the fitting outcomes we consider to have failed. Note, the arrow colours correspond to the final S\'ersic index values only (if it is grey then the S\'ersic index is outside the range of the values we considered physical. see Sec.\ref{sec:reject}) and each grey point has several arrows associated with it due to the combination of starting values of our initial grid. See the text for a detailed description.}
\label{fig:converge}
\end{figure*}

\noindent
{\bf (4) Reversal of the bulge and disc components}

\noindent On occasion the initially assigned bulge component
migrates to fit the disc component and the disc to the bulge.
This effect was first noted in \cite{Allen2006} and can be 
rectified by switching the components if necessary. Here regardless of the initial
parameters we assign the component with the lowest half-light radius
as the bulge (i.e., inner, more compact component) and the other as the disc 
(i.e., outer, more extended component).
This can, however, lead to cases where the bulge has a lower S\'ersic index than the disc 
(see Appendix \ref{App:flags} for our treatment of these cases), which in the majority of cases
is an unphysical solution.

~

\noindent
{\bf (5) Default \galfit{} errors do not reflect the full complexity
  and uncertainty in the final fits. }

\noindent
It is known that \galfit{} (like other fitting codes) often underestimates
the error on the returned parameters \citep[see e.g.~][]{Haussler2007}, possibly due to
the poor treatment of correlated noise in real images. 
Essentially the final errors do not provide any indication of fit confidence.
By running \galfit{} multiple times from a grid of initial conditions we
can assess the level of convergence which can be used to provide  more realistic error estimates.
This reassessment of the errors is probably the most important outcome of our adoption of a grid
of initial conditions, providing some certainty for each galaxy as to the robustness of the fit.\\

The five strategies above proved critical in reducing the catastrophic
error rate (as assessed from visual inspection) from $\sim 20\%$ to $\sim5\%$ enabling us to dispense with the need for
a logical filter, and most importantly obtain realistic errors.

We recognise that many of the above could also be addressed by improving
the minimisation algorithm and implementing an MCMC approach which fully samples the prior distribution.
At the present time, however, in the absence of a known prior distribution and limited computing time, 
we believe our strategies minimise the obvious systematic issues which arise when using 
the \galfit 3 engine.

\subsection{Construction of a robust decomposition catalogue}
 
\subsubsection{The initial grid and convergence}
\label{sec:grid}

As stated, for completeness, we perform both single and double (\bd decomposition) 
component fits in the \textit{r}-band on all 7506 galaxies in our sample using one or 
two S\'ersic functions, respectively. We do not constrain any fitting parameters, except for the inbuilt limits within \galfit. Hence in our two component fits the bulge and disc S\'ersic indices are not set to any particular value (e.g., 1 and 4) as is often done in other studies.
We use the Structural Investigation of Galaxies via Model Analysis (\textsc{sigma}, \citealt{Kelvin2012}) wrapper code for \galfit{}
\citep{Peng2010}. As a front-end wrapper \textsc{sigma} creates cutouts from the GAMA 
regions, does a local background subtraction and detects objects and stars using 
\textsc{sextractor} \citep{Bertin1996}. To obtain reliable S\'ersic fits it is important that local background sky variations are accounted for, yet it is also important to not over-subtract light from the galaxy itself as this will lead to systematic errors in the galaxy flux measurements. Our local background subtraction is in addition to the 
background subtraction applied during mosaicing of the GAMA data. The grid size used during this additional sky estimation depends on the size of the galaxy and varies from $32 \times 32$ to $128 \times 128$ pixels.
Using this variable mesh approach was found to be the most robust method to remove small-scale sky variations without removing light from the galaxy \citep[for further details see][]{Kelvin2012}.
After the sky subtraction, \textsc{sigma} constructs a PSF using  \textsc{psfextractor} \citep{BertinPSF} which is later used to convolve the \galfit{} models.
The \textsc{sextractor} outputs are also used to inform the fitting of neighbouring objects as well as provide initial starting values for the \galfit{} run \citep[for full details on
 \textsc{sigma} see][]{Kelvin2012}.
During the actual \galfit{} routine the primary and all secondary objects are simultaneously 
modeled using a S\'ersic function. In the case of a two-component fit, \galfit{} minimises the \x over 
two S\'ersic functions centred on the primary object while also fitting the secondary objects with a 
single S\'ersic profile. \\

To identify convergence to the global minimum we use a 
grid of initial starting points (as previously discussed) for both the bulge and disc components
as described below.

\begin{itemize}
\item  Two component fitting: a total of 88 starting combinations varying input parameters as follows,
			\begin{itemize}
			\item ratio of bulge to disc size (bulge size $\slash$ disc size, R$_{SE}$= \textsc{sextractor}radius):\\
			1:1 	($R_{SE} \slash R_{SE}$),\\ 
			1:2	($0.75 \times R_{SE} \slash 1.5\times R_{SE}$),\\ 
			1:4	($0.5 \times R_{SE} \slash 2 \times R_{SE}$),\\ 
			1:9	($0.33 \times R_{SE} \slash 3 \times R_{SE}$)
			
			\item two sets of component starting S\'ersic index:\\  
			 n=4+1 (bulge + disc) \\
			 n=2.5+0.7 (bulge + disc) 
			
			\item component bulge and disc flux ratio:\\ 60\%:40\% (bulge : disc) to 10\%:90\% in steps of 5\%\\
\end{itemize}			 

\item Single component fitting: a total of 33 combinations of the input parameters
\begin{itemize}
\item $R = 2, 1, 0.5, 0.25 \times R_{SE}$
\item S\'ersic index $n= 1, 2, 3, 4$
\item total magnitude $mag= 1, 0.8 \times M_{SE}$
\item 1 additional model starting with $R=R_{SE}$ and $mag=M_{SE}$ and $n=2.5$
\end{itemize}
R$_{SE}$ and M$_{SE}$ denote the initial size and magnitude values taken from the \textsc{sextractor} outputs for the entire galaxy.\\
\end{itemize}

Figure \ref{fig:converge} shows an example convergence plot for a single 
component fit (top, G7848) and a double component 
fit (bottom, G250228). The plots show the grid of initial conditions (grey points) 
and vectors pointing to the final solution for each parameter combination.
We plot the size versus magnitude plane for the single fits
and the size versus component light fraction plane for the double fits 
(bulge component, left, disc component, right). The colour bar at the top shows 
the S\'ersic indices considered and spans the same range for all convergence plots.
The arrows pointing to the final output parameters are coloured according to the final S\'ersic index. 
In practice (e.g.~Fig.~\ref{fig:converge}, bottom) not all fits converge to a plausible 
solution and hence screening is required to remove obvious bad fits. 
Dashed lines indicate fitting outcomes which were excluded due to bad values (see the screening descriptions below) or a large reduced \x.
If the (dashed) lines are grey, the final S\'ersic index was outside the range displayed in the colour bar.
The green ellipse shows the median solution and its size corresponds to the
adopted error on the median, i.e.~the error is symmetrical and taken to be the average of the 16$^{th}$ and 84$^{th}$ percentile range.
We produce convergence plots for all 1 and 2 component fits of our 7506 galaxies. 
As mentioned previously convergence towards a tight median value is by no means 
assured and a number of situations need to be managed, including: component flipping, 
unphysical solutions, and poor quality fits. We refer to this management as screening 
and define the various steps below.
 
\subsubsection{Screening via profile switching}
\label{sec:switch}

Each of our {2247} two-component systems will have 88 model outputs from our
grid of initial parameters. When fitting the galaxies component 1 has been 
assigned the bulge initial parameters and component 2 the disc initial values.
Since \galfit{} components can migrate significantly we ensure, after the fitting has finished, 
that the more compact component is taken as the bulge and the more extended as the disc. 
However, we find some cases where, even though the bulge is smaller in size, the disc has the 
higher S\'ersic index. 
visually inspecting a number of the resulting profiles we find that the more extreme cases typically are
bad fits and flag these (see Appendix \ref{App:flags}). Additionally, we relax the criterion for
switching the components and allow bulges with lower S\'ersic index than the disc if they are  
no more then 10$\%$ larger than the disc.
For 1447 galaxies at least one of the 88 parameter combinations required switching the profiles output by \galfit.

\begin{figure}
\centering
\includegraphics[width=0.45\textwidth]{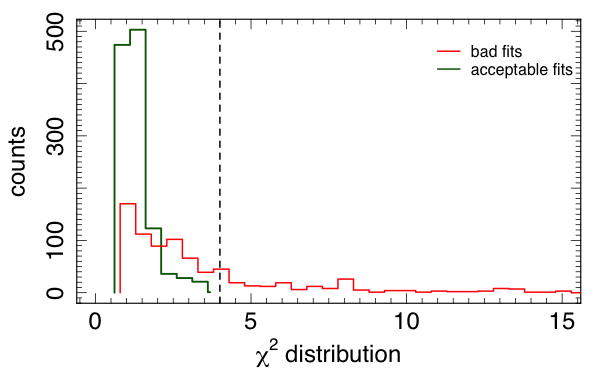}
\caption{Shown is the reduced \x{} distribution of a test sample of 
100 two-component galaxies for which 20 fitting outcomes each were visually inspected. 
The green histogram shows the fits classified as `good' and the red histogram shows the bad fits. 
The dashed vertical line is the implemented reduced $\chi~^2$ cut.}
\label{fig:chicuts}
\end{figure}

\subsubsection{Screening via rejection of poor quality fits}
We also reject fits with poor reduced \x{} values.
To decide on an appropriate reduced \x{} cut we randomly inspected 20 fitting outcomes
 for each of 100 2-component galaxies. For each fitting outcome
 we decided (by eye) whether it was acceptable or not based on the
light-profile of the model and the resulting residuals. Figure
\ref{fig:chicuts} shows the distribution of  reduced \x{} for these galaxies split into `good' and
`bad' fitting outcomes, shown as green and red histograms respectively.  The
vertical dashed line indicates the final cut of reduced \x{}$=4$ which is left
deliberately high to ensure that even for galaxies with a lot of
structure we exclude none of the acceptable fit outcomes and have enough 
outputs to evaluate the `best' fit. This cut is also implemented for the single S\'ersic fits. 
In total 20505  ($\sim 10\%$) of the 
197736 fitting results were removed from the two-component sample and 5686 ($\sim 7\%$) 
 from the 173547 fitting results of the single component sample .\\

\subsubsection{Screening via rejection of unphysical fits}
\label{sec:reject}
\begin{figure}
\centering
\includegraphics[width=0.42\textwidth]{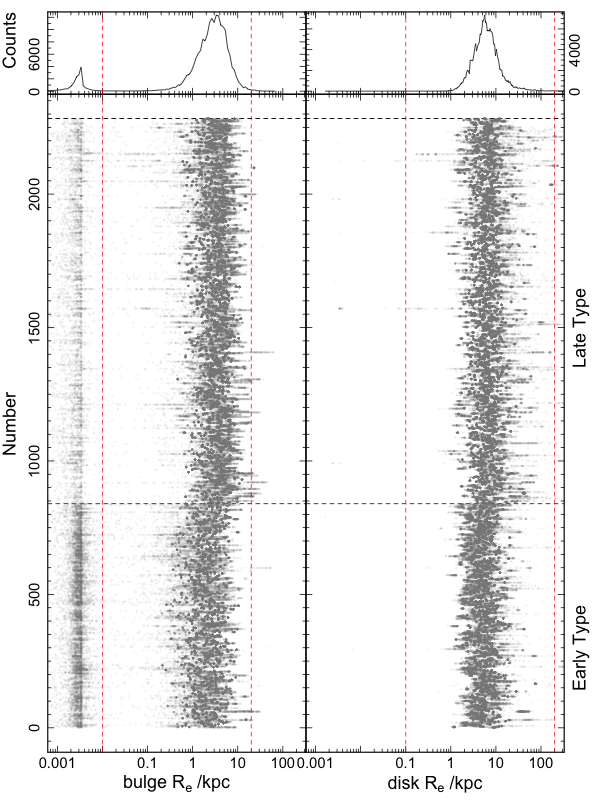}
\includegraphics[width=0.42\textwidth]{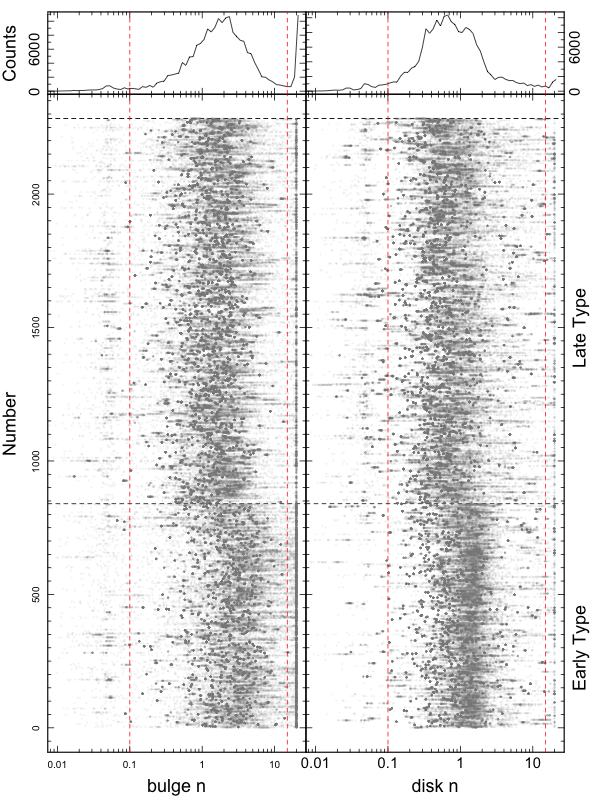}
\captionof{figure}{The top panel shows the distribution of the output size for all bulges (left) and discs (right) for the fitted two-component models (no reduced \x cut has been imposed) of each galaxy (indicated as ``Number''). The dashed vertical lines show the implemented cuts on the size before the median is established.
 The bottom panel shows the corresponding distribution for the output S\'ersic index for the bulges (left) and discs (right). In all panels the galaxies are sorted by Hubble type with the late-type 2-component systems at the top and early-type 2-component systems at the bottom. The horizontal dashed black line shows where the Hubble type changes. }
\label{fig:cuts}
\end{figure}

Fig.~\ref{fig:cuts} summarises the derived \galfit{} fitted values of all combinations for our bulges
and discs showing in the upper panels the bulge (left) and disc (right)
sizes, and in the lower panels the bulge (left) and disc (right)
S\'ersic indices. Taking the top left panel we see that the bulge
sizes follow two distinctive bands, one at plausible sizes (i.e.,
0.1-10$''$ scales), and one at unphysical sizes (0.001-0.01$''$) given
the data resolution of $\sim 1.5''$. 
We reject the fitting results which result in overly compact ``bulges'' and 
remove these from further considerations (red dashed line). Overly large bulges or
discs are not a prominent problem but we remove obvious outliers based on the
distribution of all solutions.
Similarly, in Fig.~\ref{fig:cuts} (bottom) we show the S\'ersic
index distribution. Once again
the vertical red dashed lines indicate the division between the fitting results we
consider physical and those we consider unphysical and that should therefore
be rejected. The limits adopted leading to rejection (red dashed lines) are:\\

\begin{itemize}
\item for bulge sizes: $\rm R_e < 0.01''$ or $\rm R_e > 20'' $
\item for disc sizes: $\rm R_e < 0.1''$ or $\rm R_e > 200''$
\item for bulge and disc S\'ersic index: $n<0.1$ or $n > 15$
\end{itemize}

These cuts are deliberately permissive and should cut out only the
most unrealistic fitting outcomes.
For our 2-component galaxy sample, of the 
197736 combinations fitted, we reject 17343 ($\sim 18\%$) 
based on bulge size, 616 based on disc size ($< 1\%$), 18357 ($\sim 19\%$) based on bulge 
S\'erisc index and 11462 ($\sim 12\%$) based on disc S\'ersic index.
For the single component fits we reject fitting outcomes based on the same limits 
as the disc size and S\'ersic cuts of the 2-component fits.
Of the 173547 combinations fit to the single component systems, we reject 2753 ($\sim 3\%$)  
based on their size and 2355 based on their S\'ersic index ($\sim 3\%$).
In total 49688 ($\sim 25\%$) fitting results are rejected from our 2-component sample fits 
and 7757 ($\sim 4\%$) from our single component fits. Note, in many 
cases fitting results are rejected by more than one criterion 
(i.e. reduced \x and/or size and/or S\'ersic index).\\

We also screen our galaxies for various flags, described in Appendix \ref{App:flags}.
However, we only consider two flags important during the component \msr{} relation fits, namely
the very high (or low) \bt{} galaxies and reversed S\'ersic index galaxies.
We deem the high (and low) \bt{} galaxies single component systems and move
them from our 2-component sample to our single component sample.
Galaxies with inverted S\'ersic index have bulges with lower n than discs.
Visually inspecting several of the profiles we find that in most cases these are bad fits, i.e.,
we find the disc S\'ersic index $n>2$. This itself would not be a problem 
if the errors reflect our confidence in the fit. Many of these profiles, however,
have converged to this unphysical solution.  We find 182 late-type 2-component systems
and 87 early-type 2-component systems have inverted S\'ersic index and converged profiles.
We remove these galaxies from our component consideration, but use their single component
profile fits to establish their global \msr{} relation.

\subsubsection{Final parameter selection}
\label{subsub:parmselect}

For each galaxy we consider two possible profile fit solutions taken from the remaining fitting results:
\begin{enumerate}
\item the minimum $\chi^2$ model with the associated GALFIT parameters and errors, and
\item  the median fit values of the remaining fitting results and the 16$^{th}$ and 84$^{th}$ percentiles (i.e. the 1 $\sigma$ deviation of a normal distribution)  as an uncertainty indicator.
\end{enumerate}

While the minimum \x{} solution should represent the best formal fit from our grid,
the median model is our preferred solution, as the errors on the median reflect 
the level of convergence and robust errors are critical.
Note that the median values are calculated for each output
parameter individually and do not directly represent any single solution.  
In cases where the fitting converged, the median and minimum \x solution
will be almost identical and the 16$^{th}$ and 84$^{th}$ percentile range often
is smaller than the \galfit{} errors. We therefore adopt an error floor of 10$\%$
of the median value, which assures that in almost all cases the median solution 
is consistent with the minimum \x{} solution within the estimated errors.

\begin{figure}
\centering
\includegraphics[width=.45\textwidth]{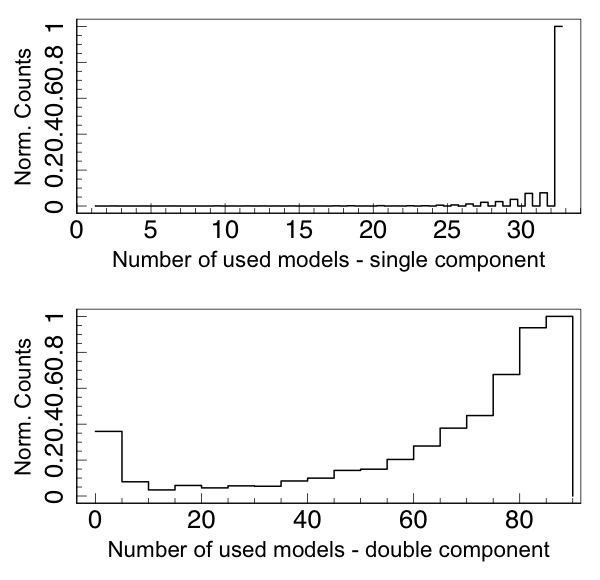}
\captionof{figure}{Here we show the distribution of the number of fitting outcomes used to calculate the median.
The top panel shows the distribution for single component fits based on single component galaxies only. The bottom plot shows the distribution of the number of 2-component fits used for 2-component galaxies only.
It can be seen that the single S\'ersic fits generally converge nicely and the 2-component fits have a broader distribution with a spike at very low numbers.}
\label{fig:models_used}
\end{figure}

Fig.~\ref{fig:models_used} shows the histogram of the number of the remaining fitting results 
used to calculate the median for all single component systems (top) and all 2-component systems (bottom). 
The single component fits often converge and
the histogram peaks at $\sim$33 fitting results with only a small tail towards lower numbers.
The 2-component fits on the other hand do not converge as often.
It is encouraging that the peak is  $\geq$85 fitting results, however, there is a large
fraction of galaxies for which very few solutions remain for the median calculations. 
In addition, there is a rise towards very low
numbers indicating that some galaxies are likely too complex to be fit with two components only.  
We find that, while the galaxies with low model counts span the whole mass range, most of them 
lie close to our upper redshift boundary and were classified as late-type double component systems. 
This shows the inherent difficulty of fitting multiple component systems in poorer image quality regimes.
In addition to the tightness of the median errors, we can also use the number of fitting results 
left for the calculation of the median
to help establish our confidence in the fitting results.

\subsection{Convergence Examples}
\label{App:converge_example}

We finish this section by presenting five examples which highlight some of the issues encountered and show that the median values present a robust alternative to the minimum \x{} solutions.
Figure \ref{fig:converge2} (upper panels) shows the convergence plots
(as introduced in Section \ref{sec:grid} and Fig.~\ref{fig:converge}), and diagnostic plots for the median and minimum reduced \x{} solutions (middle and lower panels, respectively). The 4 images which make up the diagnostic plots (middle and lower left panels) show, 
from the top left in a clockwise direction, the SDSS \textit{r}-band image stamp, the model produced by \galfit, the residual and
the \textsc{sextractor} segmentation map overplotted on the SDSS image stamp (the primary object is shown in purple and the secondaries in green). 
The red and blue ellipse show the R$_e$ of the bulge and disc, respectively. 
The yellow ellipse is the original \textsc{sextractor} radius and the cyan ellipses show the radii for which the surface brightness was evaluated. Also shown (middle and lower right panels) are the 1D light profile comparisons. 
The black points are the values extracted from within the blue ellipses, the red and green lines are the 1D light profile for the bulge and disc, evaluated from the \galfit{} model and the green line is the total light profile.
The lower inset panel shows the residual between the model and data.
Below we discuss five examples of various fit and convergence outcomes. We only discuss examples for two-component galaxies here since we find that the single S\'ersic fits generally converge well (e.g., example a):

\begin{figure*}
\centering
\includegraphics[height=0.9\textheight]{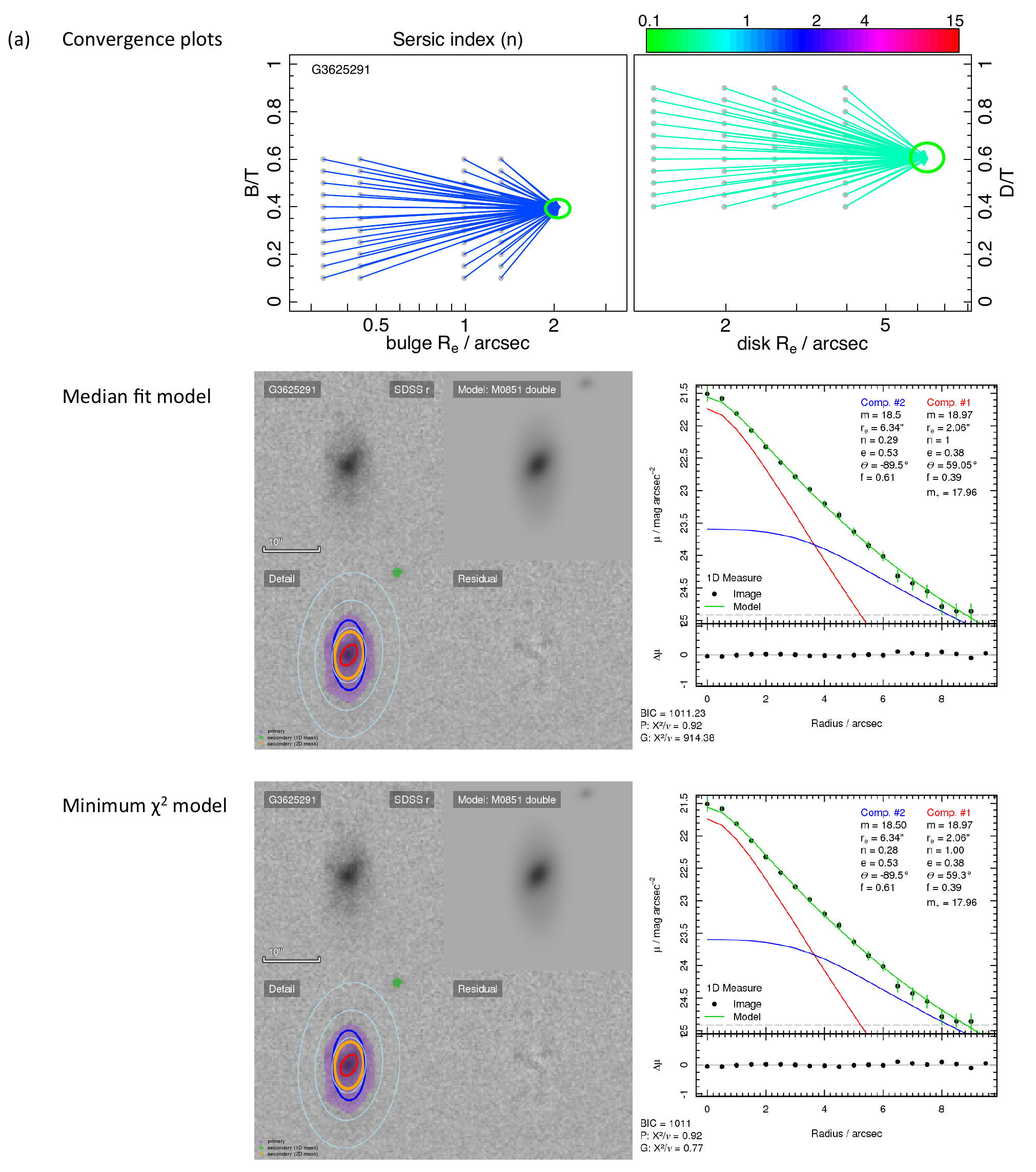}
\captionof{figure}{Presented are the convergence plots (top) and corresponding diagnostic fit plots for the median fit model (middle) and minimum reduced \x{} solution (bottom). We show four examples ranging from full convergence to no convergence (panels a, b, c, d) and 1 example where all fits are unrealistic (panel e). A detailed discussion can be found in the text.}
\label{fig:converge2}
\end{figure*}

\renewcommand{\thefigure}{\arabic{figure} (Cont.)}
\addtocounter{figure}{-1}

\begin{figure*}
\centering
\includegraphics[height=0.9\textheight]{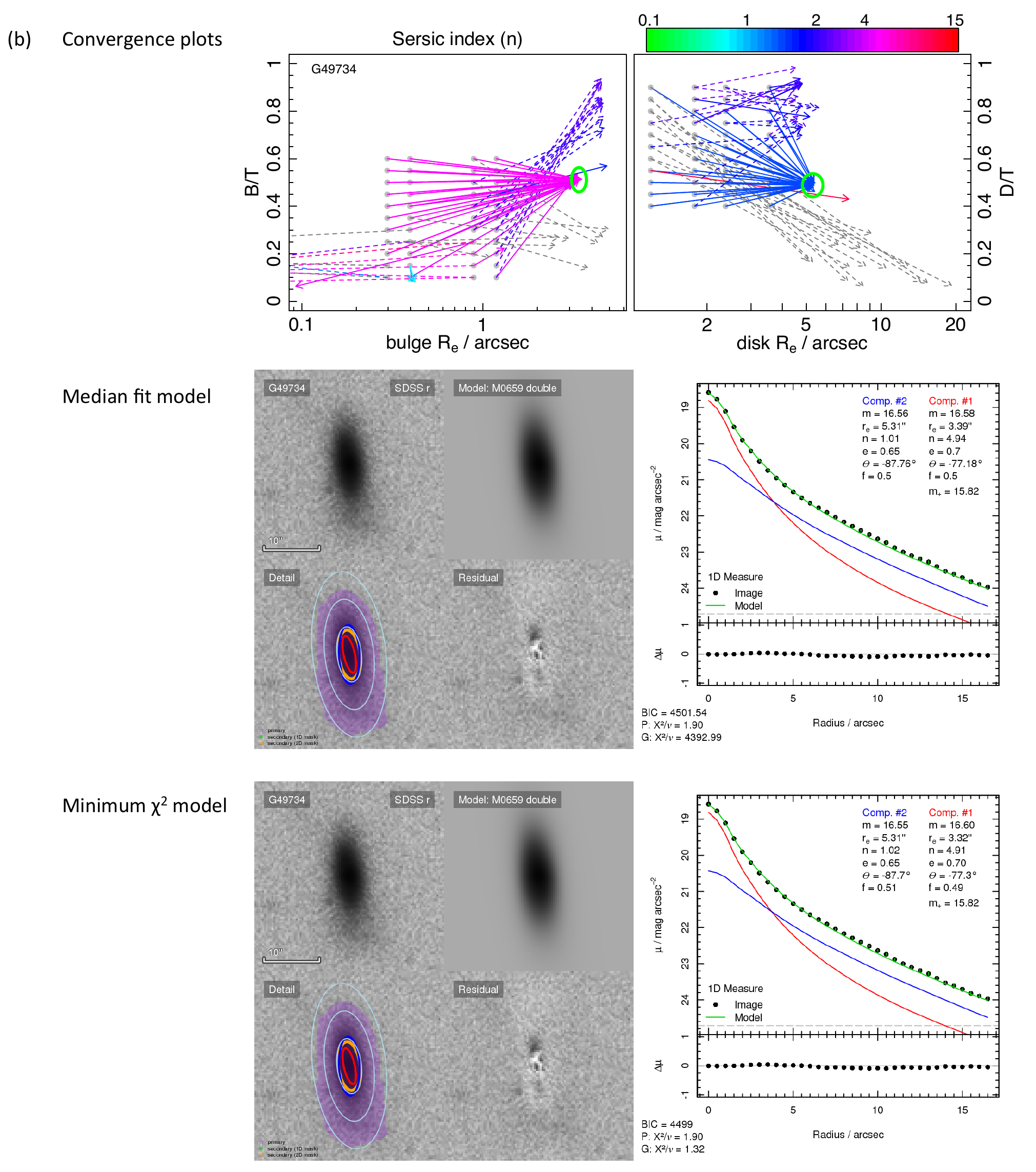}
\caption{}
\end{figure*}

\addtocounter{figure}{-1}
\begin{figure*}
\centering
\includegraphics[height=0.9\textheight]{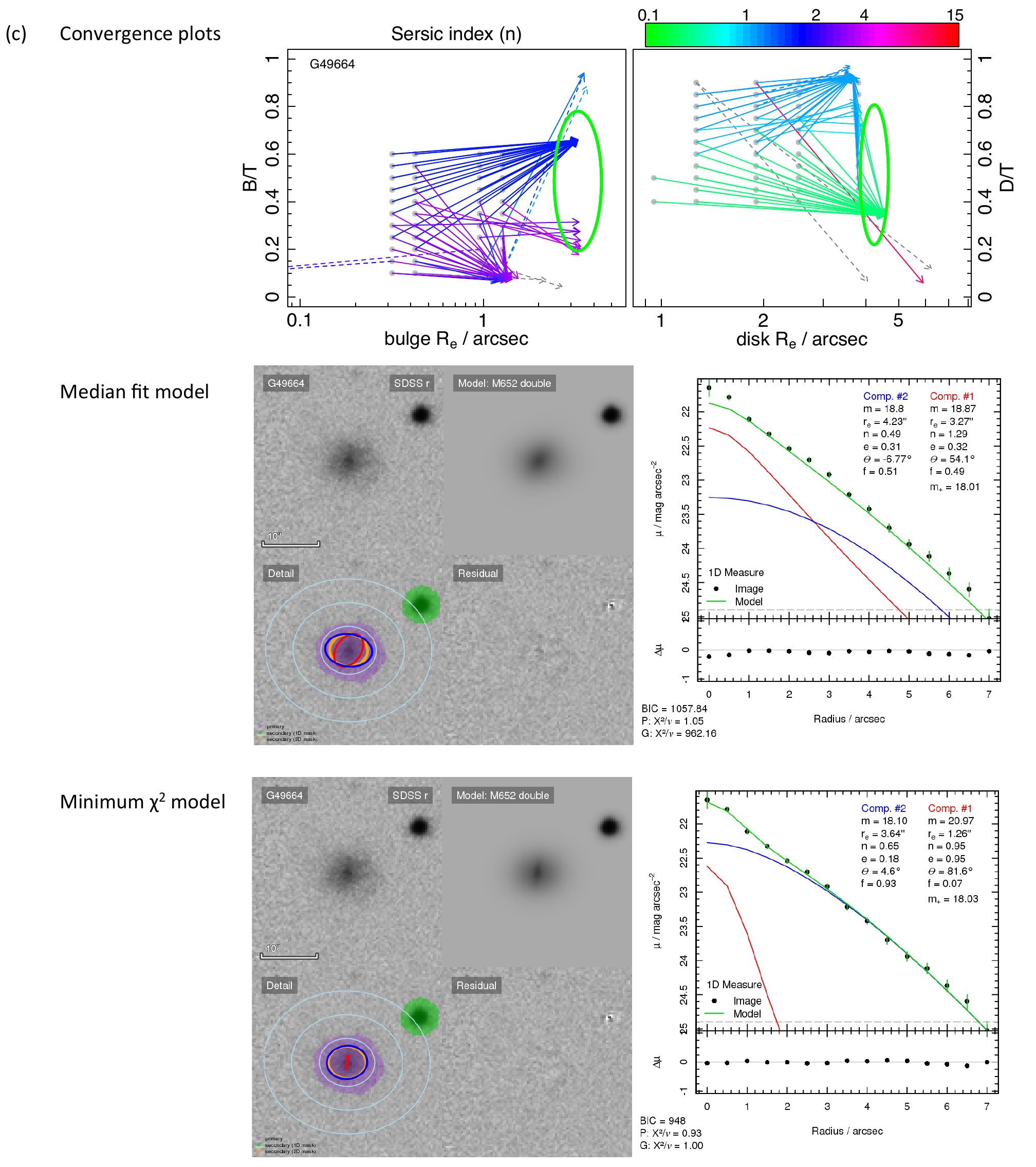}
\caption{}
\end{figure*}

\addtocounter{figure}{-1}
\begin{figure*}
\centering
\includegraphics[height=0.9\textheight]{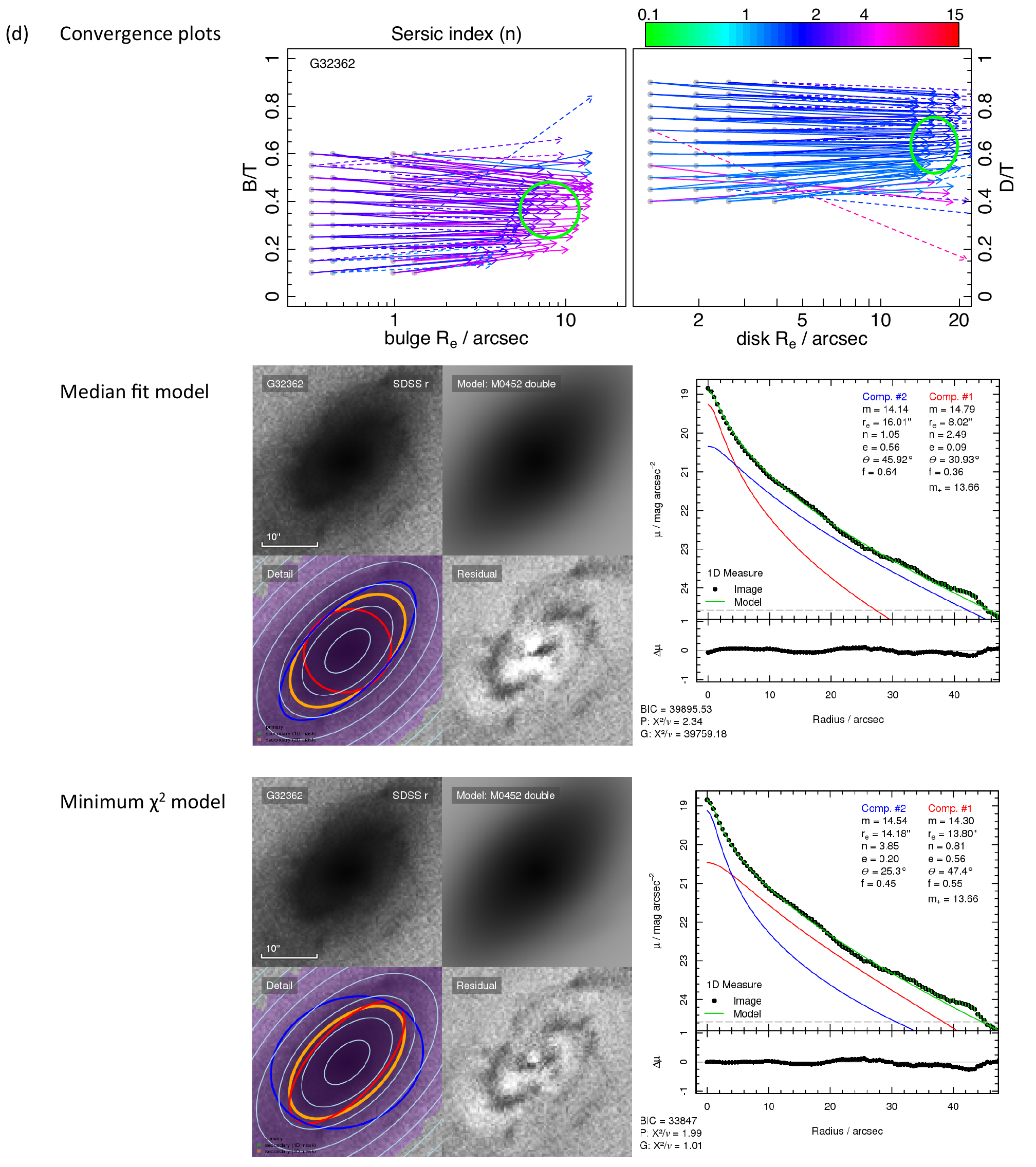}
\caption{}
\end{figure*}

\addtocounter{figure}{-1}
\begin{figure*}
\centering
\includegraphics[height=0.9\textheight]{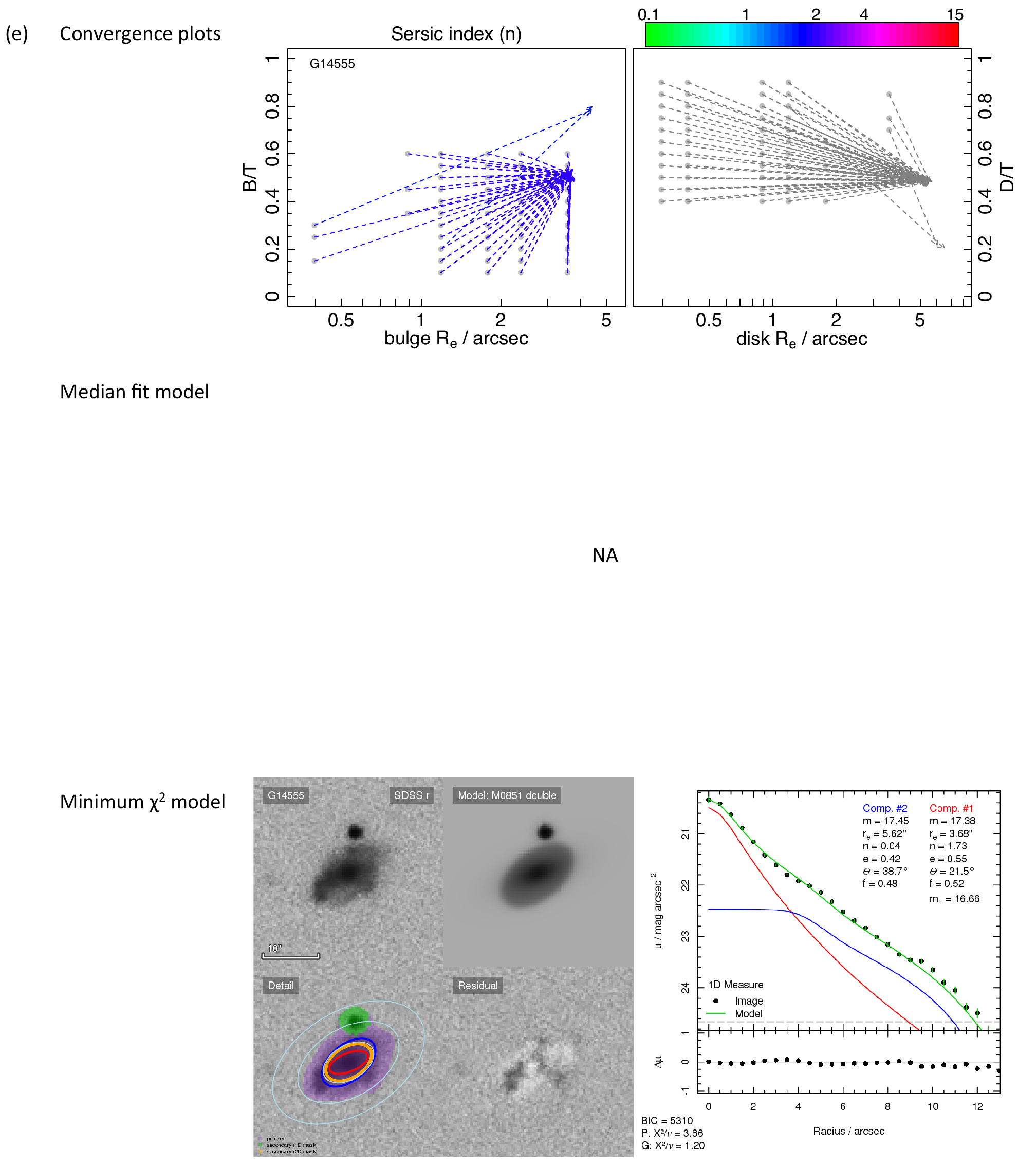}
\caption{}
\end{figure*}
\renewcommand{\thefigure}{\arabic{figure}}

\begin{itemize}
\item \textbf{Example a: full convergence}\\
Fig.~\ref{fig:converge2}a shows the ideal case of full convergence from all combinations of initial parameters. The median and minimum \x{} solution diagnostic plots also show that both reached the same answer. The residual images show little structure and the final errors of the 
median fit are small as indicated by the green ellipse on the convergence plot.
For 2-component fits we consider them fully converged when they have more than 80 fitting results remaining
after rejection of spurious fits and the error on the
median is set to the 10$\%$ error floor. This is the case for 423 ($\sim 19\%$) galaxies.
Similarly, for single components over 30 fitting results must remain for the median calculation and all errors are set to the 10$\%$ error floor.
This is true for 4566 ($\sim 87\%$) single component galaxies. \\

\item \textbf{Example b: partial convergence}\\
The median and minimum reduced \x{} model diagnostic plots in Fig.~\ref{fig:converge2}b show good agreement. From the convergence 
plot it is obvious that many of the solutions found by 
\galfit{} were rejected during the screening process, due to an unphysical S\'ersic index or high reduced \x.
The remaining models after screening show convergence resulting in a good solution with tight errors.
For 2-component systems we consider good convergence to be reached when we have 60 to 80 solutions remaining, with the errors set to the 10$\%$ error floor. This is the case for 297 ($\sim 13\%$) galaxies. 
Equivalently, for single components 25 to 30 solutions must remain for the median calculation with the errors set to the error floor. We find this true for 220 ($\sim 4\%$) single component galaxies. \\

\item \textbf{Example c: two plausible solutions} \\
The diagnostic plots in Fig.~\ref{fig:converge2}c suggest that the median model gives a physically more meaningful 
two-component solution than the minimum reduced \x solution which is converging towards a single 
component solution. The convergence plot, however, highlights that the median solution is not 
one of the actual solutions found by \galfit. Nevertheless, the errors on the median 
(green error ellipse) enclose both solutions. While the median fit can not be considered
as robust this uncertainty is fairly reflected in the final errors.
To establish whether several plausible solutions have been found, we test how many solutions are near the median.
If less than 10$\%$ of the solutions of the median for at least one of the
size, S\'ersic index or \bt{} values of either the bulge or disc lie within 10$\%$ (i.e. the error floor) 
then we consider the fits to have converged to several plausible solutions which are distinct from the median. 
This is the case for 205 ($\sim 5\%$) double component systems. For the single component systems
we consider the size, S\'ersic index and magnitude and find 6 ($<1\%$) single component galaxies 
have converged to several distinct solutions.\\

\item \textbf{Example d: no convergence}\\
Fig.~\ref{fig:converge2}d shows a case where no obvious single converged solution is found, but the median model returns acceptable parameters with an appropriately broad error distribution. The diagnostic plots also show that the median model returns a physically possible solution with good residuals. The minimum \x{} solution, however, 
returns a fit where the bulge, even though it has a smaller R$_e$ is the dominant component in the outer parts of the galaxy. Since the
errors associated with the median model are large this particular galaxy will not have much influence on the \msr{} relation we fit in Sec.~\ref{sec:strucMSR}, but using the median parameters and large
error bars means that the galaxy will not be discarded from the sample.
To test non-convergence we use the same metric as in example 3, i.e. the percentage of solutions found within 10$\%$ of the median.
We consider galaxies not clearly converged if more than 10$\%$ but less than 50$\%$ of the solutions lie close to the median.
We test the size, S\'ersic index and \bt{} measurements for the bulge and disc and find for 931 ($\sim 41\%$) 
2-component systems at least one of them is not converged. For the single component systems this is the case 
for 141 ($\sim 3\%$) galaxies.\\

\item \textbf{Example e: no solution}\\
Fig.~\ref{fig:converge2}e shows a case where convergence is found, however, all fits are excluded from the final catalogue due to the screening process. No median model diagnostic plot is shown due to all fit parameters being unrealistic. Only 120 ($\sim 5\%$) of our 2-component galaxies and 129 ($\sim 2.5\%$) of our single component systems fall into this category.
\end{itemize}

Convergence plots for all systems are available from the GAMA
database.

\begin{figure}
\centering
\includegraphics[width=0.45\textwidth]{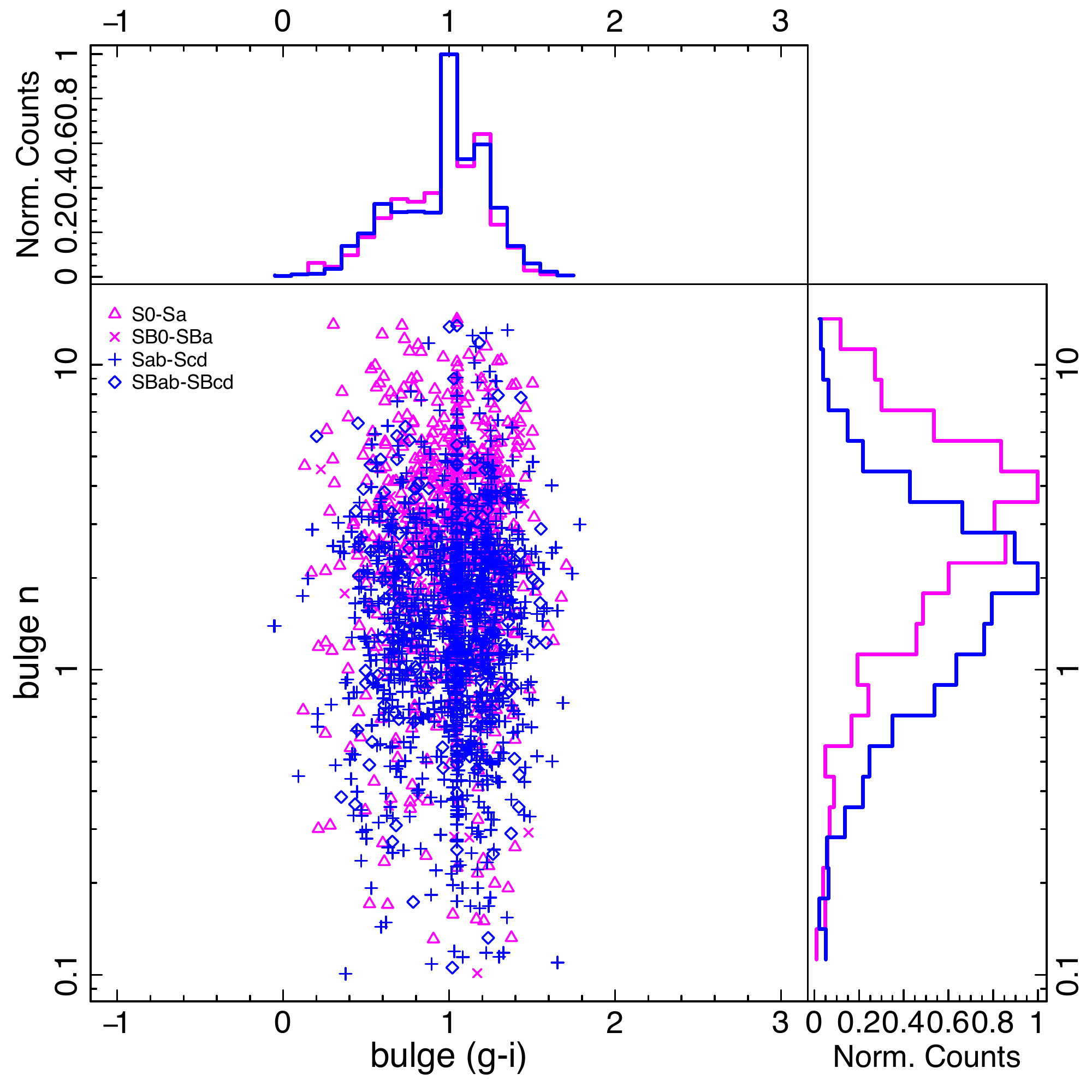}
\includegraphics[width=0.45\textwidth]{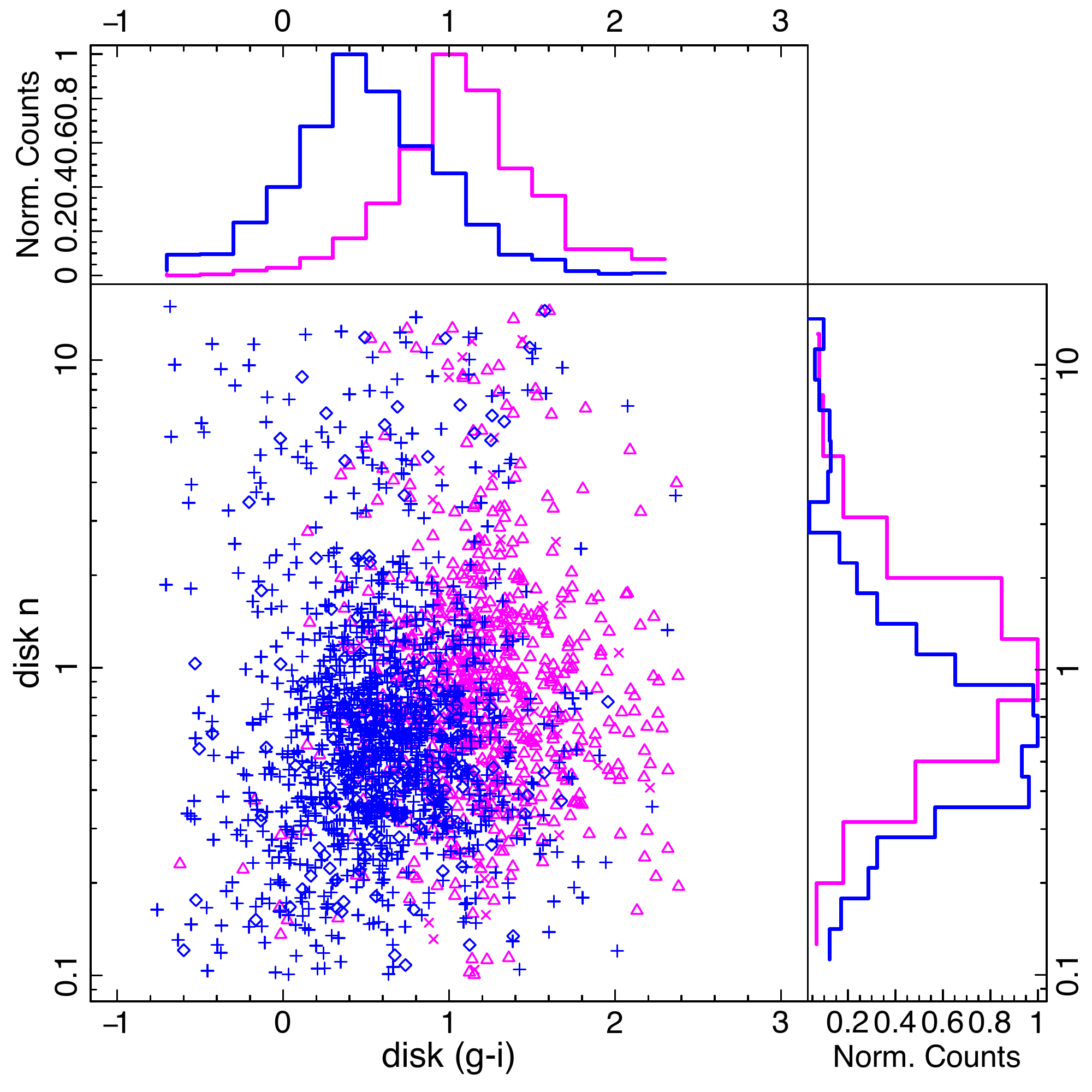}
\caption{Shown are the (\textit{g-i}) colour vs S\'ersic index distributions 
for the bulges (top) and discs (bottom) in our two-component galaxy sample. 
The colour coding in both plots is the same with late-types in blue and 
early-types in magenta.
The bulges of late-type galaxies have smaller
 S\'ersic indices than the early-type bulges. The S\'ersic index distribution 
of the late-type discs also peaks slightly lower than the early-type discs. 
Late-type discs are also bluer than early-type discs.}
\label{fig:colourn}
\end{figure}

\newpage
\section{Component masses}
\label{sec:compmass}

To derive the \msr{} relations for our galaxy components we now
need component mass estimates.
For the single S\'ersic fits we can directly use the GAMA stellar mass estimates 
from the StellarMassesv18 catalogue and apply the fluxscale correction 
\citep[for a detailed description see][essentially the fluxscale correction accounts for 
the differences between aperture matched and S\'ersic photometry]{Taylor2011}. 
These masses are based on synthetic 
stellar population models from the BC03 library \citep{BC03} with a
\citet{Chabrier2003} initial mass function and the \citet{Calzetti2000}
dust obscuration law. We find that for our sample the typical error, which
has been derived in a Bayesian way \citep[Sections 3.2-3.4 of][]{Taylor2011}, is of the order of $\sim 0.12$ dex.\\

For the double component galaxies we calculate the component mass from the 
component colours \citep{Driver2006} using the relationship between
optical colour ($g-i$) and mass-to-light ratio as calibrated by \cite{Taylor2011}: 
\begin{equation}
\label{eq:mass}
\log{M_{*}/M_{\odot}} = -0.68 + 0.7~(g-i) - 0.4~(M_{i}-4.58)
\end{equation}
where M$_i$ is the absolute magnitude in the \textit{i}-band 
and we use the  (\textit{g-i}) colour of either the bulge or disc to calculate the component mass.
The stellar masses derived via Eq.~\ref{eq:mass} are estimated to be accurate within a factor of two.
Note that this equation is sensitive to the evolution of colour and magnitude, however, as our sample 
has a low redshift range the effects will be negligible. 

Ideally we would use bulge and disc colours derived from the \bd{} decompositions in the
\textit{g}- and \textit{i}-band, however, this is beyond the scope of this paper.
Hence we have to estimate the colours of the components. For this we measure
the PSF and total magnitudes of our galaxies in the \textit{g, r,} and \textit{i}-band and we then
use the \galfit{} measured \textit{r}-band component magnitudes and \bt{} to estimate the 
bulge and disc colours. 

We measure the core (i.e. PSF) and total magnitudes, which we correct for foreground extinction, 
in the \textit{g, r,} and \textit{i}-band using \textsc{lambdar}, a code developed to measure PSF 
weighted aperture photometry  \citep[for more details see][]{Wright2016}.
We then equate the colours measured using the PSF magnitudes to bulge colour measurements 
(i.e., assuming the bulge has no colour gradient) 
and combine these colours with our $r$-band bulge magnitude from \galfit{} to
obtain bulge flux measurements in both $g$ and $i$ (i.e.,
$m_{i,bulge}=m_{r,bulge} - (r-i)_{PSF,bulge}$). In cases where the bulge
colours could not be measured, we use the median bulge colour of the
entire population as there is no significant trend between bulge colour and mass. 
We derive $g$ and $i$ disc fluxes by assuming that the disc 
flux in each band is equal to the \textsc{lambdar}
total flux minus the previously derived bulge flux. 

We examine the disc ($g-i$) colour distribution and find a small 
number of extreme outliers ($>3\sigma$) from the colour
distribution, whose disc ($g-i$) colours we subsequently replace with the running
median disc ($g-i$) colour.

Finally, we use the derived bulge and disc ($g-i$) colours and total $i$-band
magnitudes to derive stellar mass estimates for each component according 
to equation \ref{eq:mass}.
The component colour versus S\'ersic index distribution is shown in Fig.~\ref{fig:colourn} 
for bulges (top) and discs (bottom).
Figure \ref{fig:colourn} shows that for late-type galaxies bulges are generally redder than discs but for
 early-type galaxies the colours are very similar
\citep[for a detailed study of the wavelength dependence of \bd decompositions in GAMA see][]{Kennedy2016}.
 But it also highlights the problem galaxies for which the component colour had to be 
 set to the median colour  (vertical band in the bulge plot, top)  in order to be able to 
 calculate a stellar mass estimate. 

\begin{figure}
\centering
\includegraphics[width=0.44\textwidth]{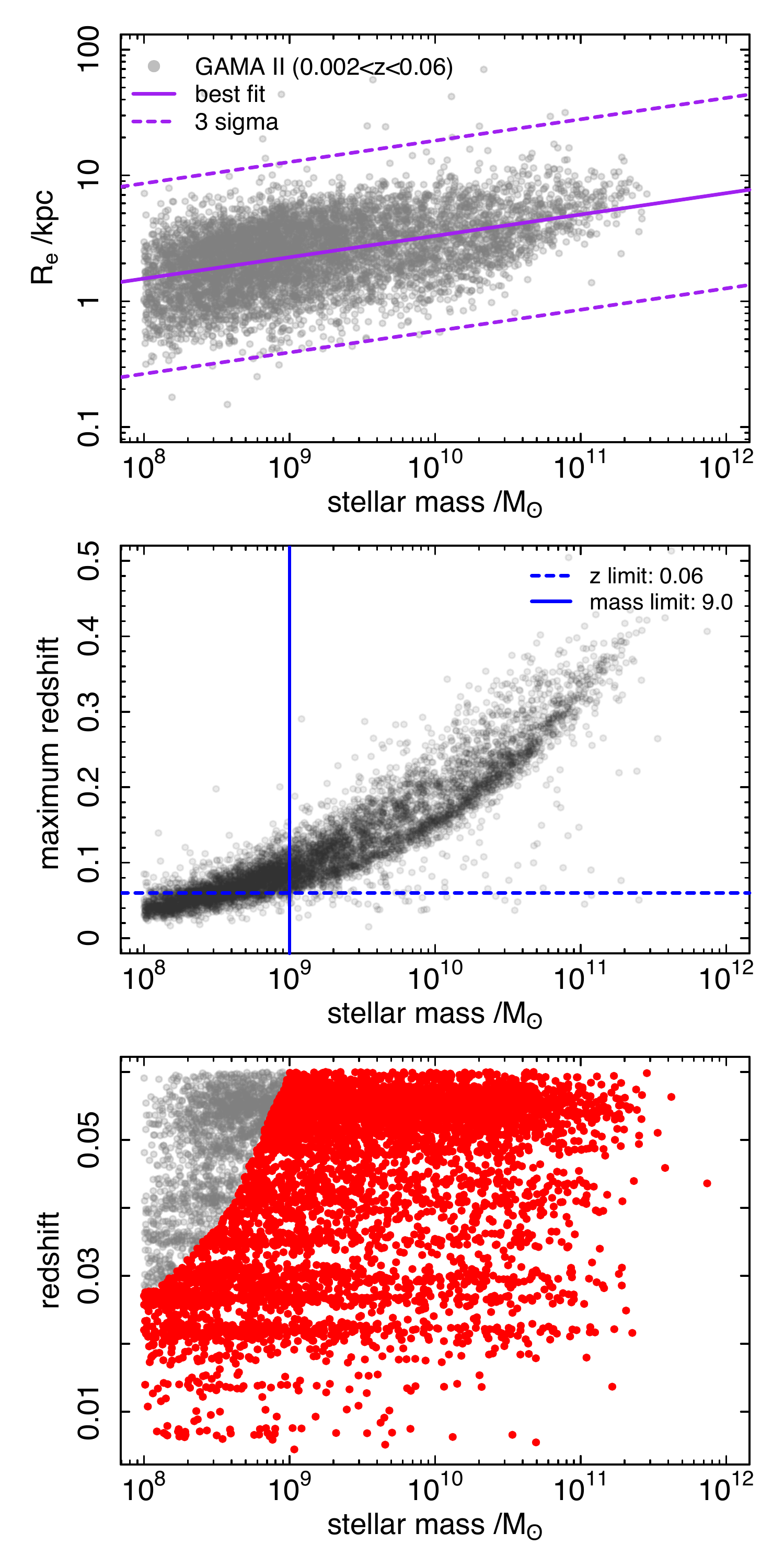}
\caption{The top panel shows the total stellar mass -- half-light size distribution (derived from single S\'ersic fits) of the GAMAnear sample. All galaxies more than 3 sigma offset from the line of best fit are removed as outliers from our \msr{} relation fits.\newline
The middle panel shows the total stellar mass - maximum redshift distribution of the sample. The blue dashed line shows our redshift limit and the solid blue line the lower mass limit for a volume limited sample. All galaxies to the right of the mass limit and above the redshift limit are included in the volume limited sample.\newline
The bottom panel shows the total stellar mass - redshift distribution of our sample. For all galaxies below $10^9\mathcal{M}_{\odot}$ we have implemented a smooth volume limited sample selection. Our final sample is highlighted in red.}
\label{fig:sample}
\end{figure}

\section{\msr{} relations}
\label{sec:msr}

We now present the \msr{} relations, firstly for the
 different Hubble types and secondly by structural components. We conclude our analysis by
 presenting a combined \msr{} relation for disc (i.e.,~Sd-Irr galaxies and disc
 components) as well as spheroids (i.e.,~ellipticals and classical bulge components).\\
 
Figure \ref{fig:sample} shows the selection of the final sample, based on total stellar mass and single S\'ersic profile fits, used to fit the \msr{} relation.
Note, we do not consider any galaxies below  $\mathcal{M}_*=10^8\mathcal{M}_{\odot}$, 
since number counts are too low to establish a robust weight (after fluxscale correction
this reduces the sample to 6788 galaxies).
First we find and exclude outliers from the general mass-size distribution (top panel of Fig.~\ref{fig:sample}). 
For this we fit the entire sample with a simple power law and remove all  galaxies which are more than 
3 sigma offset from the best fit linear relation (28 galaxies in total).
We then establish the lower mass limit for a volume limited sample at $z=0.06$. In the middle panel of
 Fig.~\ref{fig:sample} we plot the maximum redshift at which each galaxy can be seen 
 versus its stellar mass. To establish the lower mass limit of a volume limited sample 
 we find the point at which more than 95$\%$ of our galaxies could be seen at a redshift 
 of 0.06 (i.e., their maximum redshift is $z_{max}\geq0.06$, indicated as the dashed line). 
 We find a lower mass limit of $\mathcal{M}_*=10^9\mathcal{M}_{\odot}$ (solid blue line), 
which would reduce our sample size to 3679 galaxies.
To include lower mass galaxies we implement a smooth volume and 
mass limited sample for galaxies below $\mathcal{M}_*=10^9\mathcal{M}_{\odot}$.
For each galaxy we evaluate if their measured redshift is larger than their expected maximum redshift and remove them (1624 galaxies removed). The bottom panel of  Fig.~\ref{fig:sample} shows the resulting sample distribution.
All galaxies in red are included in our final sample (5136 total) and all grey points are excluded from the volume limited sample.
For all galaxies below $\mathcal{M}_*=10^9\mathcal{M}_{\odot}$ we also calculate a V/V$_{max}$ weighting based on their redshift and our sample redshift limits of $0.002<z<0.06$. For galaxies with $\mathcal{M}_*>10^9\mathcal{M}_{\odot}$ the V/V$_{max}$ is set to 1.
To ensure that the we only include galaxies with good, physical fits we require:
\begin{itemize}
\item the fluxscale correction is within 0.5 and 1.5; \\
\item the components have S\'ersic indices between 0.3$< n <$10; \\
\item the components are resolved, i.e. R$_e>0.5\times$FWHM of the PSF which is determined 
from each galaxy's fit image individually;\\
\item at least 5 solutions remained to calculate the median fit parameters.
\end{itemize} 
This reduces the sample size to 2669 single component and 1470 double component systems.\\
 
We adopt a simple power law, following \cite{Shen2003} and \cite{Lange2015}, to fit the \msr{} relation:  
\begin{equation}
\label{equ:rm1}
R_e=a \mathrm{\left(\frac{\mathcal{M}_*} {10^{10}\mathcal{M}_{\odot}}\right)}^{b},
\end{equation}
 where R$_e$ is the effective half-light radius in kpc and $\mathcal{M}_*$ is the mass of the galaxy.
To perform the actual fitting we utilise the \textsc{hyperfit} package \citep{Robotham2015}  
which estimates the \msr{} relation 
via Bayesian inference for each morphological group and component.
During fitting we assume uniform priors and each galaxy is weighted by its  V/V$_{max}$ and 
the (convergence) errors for each individual galaxy are fully taken into account during the fitting process.

\subsection{Global \msr{} relations by Hubble type}

\begin{table}
\centering
\begin{tabular}{ll|l|l}
	\hline
	\multicolumn{2}{l}{Hubble type}&&\\
(i)					&& a{ } /kpc 		& b  \\ 
	\hline \hline
Sd-Irr 		&&  6.347 $\pm$ 0.174 & 0.327 $\pm$ 0.008\\
S(B)ab-S(B)cd 		&&  5.285 $\pm$ 0.098 & 0.333 $\pm$ 0.009\\
S(B)0-S(B)a 		&&  2.574 $\pm$ 0.051 & 0.326 $\pm$ 0.015\\
E 		&&  2.114 $\pm$ 0.035 & 0.329 $\pm$ 0.01\\
LBS 		&&  2.366 $\pm$ 0.166 & 0.289 $\pm$ 0.019 \\
\hline 
	(ii)			\\
E ($M_*\geq10^{10}M_{\odot}$)		&&  1.382 $\pm$ 0.065 & 0.643 $\pm$ 0.032\\
E ($M_*\geq 2 \times 10^{10}M_{\odot}$)		&&  0.999 $\pm$ 0.089 & 0.786 $\pm$ 0.048\\
E ($M_*<10^{10}M_{\odot}$)		&&  1.978 $\pm$ 0.077 & 0.265 $\pm$ 0.022\\
E ($M_*< 2 \times 10^{10}M_{\odot}$) 		&&  2.108 $\pm$ 0.041 & 0.326 $\pm$ 0.012 \\
	\hline
	\multicolumn{4}{l}{ }\\
	\multicolumn{3}{l}{Structural Components}&\\
(iii)			\\
	\hline \hline
late type disc 		&&  6.939 $\pm$ 0.17 & 0.245 $\pm$ 0.008\\
late type bulge 		&&  4.041 $\pm$ 0.129 & 0.339 $\pm$ 0.014\\
early type disc 		&&  4.55 $\pm$ 0.097 & 0.247 $\pm$ 0.015\\
early type bulge 	&&  1.836 $\pm$ 0.054 & 0.267 $\pm$ 0.026 \\
	\hline
	\multicolumn{4}{l}{ }\\
	\multicolumn{2}{l}{Combined Case} &\\
(iv)				\\
	\hline \hline
all discs 		&&  5.56 $\pm$ 0.075 & 0.274 $\pm$ 0.004\\
all discs + LTB		&&  5.125 $\pm$ 0.065 & 0.263 $\pm$ 0.004\\
final $z=0$ discs 		&&  5.141 $\pm$ 0.063 & 0.274 $\pm$ 0.004\\
E + ETB 		&&  2.033 $\pm$ 0.028 & 0.318 $\pm$ 0.009\\
final $z=0$ spheroids		&&  2.063 $\pm$ 0.029 & 0.263 $\pm$ 0.005 \\
\\
global late-types 		&&  4.104 $\pm$ 0.044 & 0.208 $\pm$ 0.004 \\
\hline
\end{tabular}
\captionsetup{width=0.5\textwidth}
\caption{The regression fit parameters to Eq. \ref{equ:rm1}
  for the different Hubble types and structural components  
  as well as a combined early and late type
  relation (see Fig.\ref{fig:bd1}).}
\label{table:compfits}

\begin{tabular}{ll|l|l}\\
\\
	\hline
late-type							&& a{ } & b \\
	\hline \hline
r band  								&& 3.971 $\pm$ 1.745  		& 0.204 $\pm$ 0.018 \\
	\hline	
early-type	 						&& a{ } & b \\
	\hline \hline
r band 								&& 1.819 $\pm$ 1.186 		& 0.46 $\pm$ 0.023  \\
M$_*>$2$\times 10^{10}$M$_{\odot}$	&& 1.390 $\pm$ 1.557			& 0.624 $\pm$ 0.033 \\
\hline
\end{tabular}
\caption{Fitting parameters taken from L15 (Table 2, 3 and B2) for the 
morphological late- and early-type \msr{} relation.}
\label{table:L15}
\end{table}

To establish the global (i.e. single component S\'ersic fit) 
\msr{} relation by Hubble type we have grouped the GAMAnear sample 
into 5 populations:
\begin{itemize}
\item 1564 late-type single component galaxies (including 40 high/low \bt{} galaxies), 
\item 890 late-type multi-component systems (comprised of Sab-Scd and SBab-SBcd galaxies) of which 708 have also good 2-component fits,
\item 580 early-type multi-component systems (which include S0-Sa and SB0-SBa galaxies) of which 493 have also good 2-component fits,
\item 806 early-type single component (including 33 high \bt{} galaxies), and 
\item 372 Little Blue Spheroids (LBS)
\end{itemize}

The resulting global \msr{} relations  
are shown in panel (i) of Fig.~\ref{fig:bd1}, 
from left to right the plots are (a) Sd-Irr, (b) visually late-type multi-component systems, 
(c) visually early-type multi-component systems, (d) ellipticals and e) LBS.
The fit parameters can be found in Table \ref{table:compfits} (i). \\

We find that the single S\'ersic \msr{} relation fits to the different morphological types
lie on almost parallel lines (i.e., comparable gradients but offset in normalisation). Most two-component systems are more
massive than Sd-Irr galaxies, but compared at the same mass we find that two component systems are smaller than Sd-Irr galaxies. Compared to the ellipticals, however, we find that two-component systems are larger at a given mass. This corroborates the composite nature of these galaxies, i.e.~the disc surrounding the bulge makes their global R$_e$ appear larger than ellipticals but smaller than Sd-Irr galaxies at a given mass, with the offset between the Sd-Irr and elliptical relation depending on the relative dominance of the disc or bulge component. 
The LBS galaxies on the other hand are our smallest and least massive population. The slope of their \msr{} relation is flatter than that of any of the other morphological types. Nevertheless, their sizes are mostly consistent with an extension of the elliptical population.  In fact, within the errors the LBS relation is  consistent with the low-mass elliptical relation ( $\mathcal{M}_*<10^{10}\mathcal{M}_{\odot}$), see Table \ref{table:compfits} (ii).\\

Here our morphological subdivisions are finer than in our previous work (L15) but broadly agree.
In detail our Sd-Irr class has a steeper \msr{} relation than the late-type relation in L15. This is
an effect of our sample selection. In fact, fitting an \msr{} relation to a combined sample of Sd-Irr and all two-component 
systems (see Table \ref{table:compfits} (iv)) results in a relation fully consistent with the late-type relation in L15 (reproduced in Table \ref{table:L15}).   
Our elliptical class has a shallower \msr{} relation compared to the early-type relation in L15.
This is also largely a sample selection effect, caused by the relative increase
 in the number of low-mass to high-mass ellipticals within the sample.
 Fitting a high-mass elliptical relation (see Table \ref{table:compfits} (ii)), 
 similar to L15, with $\mathcal{M}_*>10^{10}\mathcal{M}_{\odot}$ we again
 find good agreement with our earlier work.

\subsection{\msr{} relations by galaxy component}
\label{sec:strucMSR}

\begin{figure*}
\centering
\includegraphics[width=0.9\textwidth]{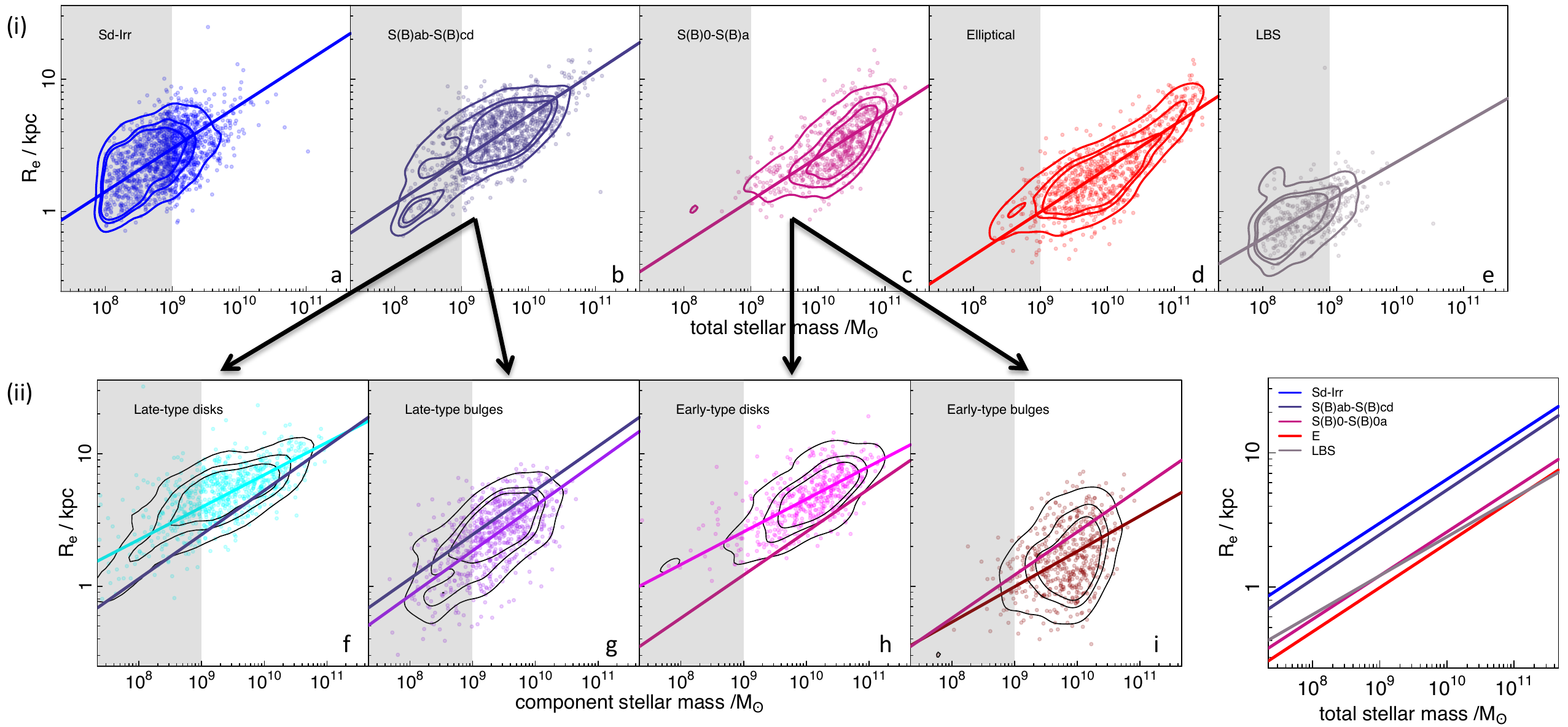}
\caption{The top panel (i) shows the global \msr{} relation for: a) Sd-Irr (blue), 
b) late-type multi-component (dark purple), c) early-type multi-component 
(rose), d) ellipticals (red), and e) LBS (grey) galaxies. \newline 
Panel (ii) shows, from left to right, the component \msr{} relation 
for late-type discs (cyan) and bulges (purple), plots f) and g) respectively. We also plot the global S(B)ab-S(B)cd relation from panel b) in comparison.
The early-type discs (magenta) and bulges (dark red) are shown in plots h) and i). Again we plot the global S(B)0-S(B)a (shown in panel c) in comparison. The  grey shaded areas indicate where our smooth volume limited sample selection starts and galaxies are up-weighted by their V/V$_{max}$. The black lines are the 90th, 68th and 50th percentiles of the respective mass--size distributions. The arrows show from which
population the components were derived.
\newline Finally we show all global \msr{} relations in comparison in the far right plot in panel (ii). This highlights the similarities between the different populations, i.e. the relations are parallel but offset from each other depending on their bulge fraction. }
\label{fig:bd1}
\end{figure*}

The distributions and fits to the structural component \msr{} relations are shown 
in panel (ii) of Fig.~\ref{fig:bd1} and the fitting 
parameters can be found in Table \ref{table:compfits} (iii).\\ 
From left to right panel (ii) shows the late-type discs and bulges 
(LTD and LTB, plots f and g, respectively)
followed by early-type discs and bulges (ETD and ETB, plots h and i). 
In each panel we also show the global \msr{} relation of the population they were derived from, 
which is indicated by the large black arrows.
Not surprisingly we find that generally discs are larger and bulges are smaller than the global 
single S\'ersic fits of the population they were derived from. 
Additionally our data shows that the component \msr{} relations are typically less curved than the global
relation they were derived from. However, the late-type discs are an exception to this and show a slight
upward curvature at high masses. This is in qualitative agreement with results from \cite{Bernardi2014} 
who found that the discs of late-type galaxies cannot be fit with a single power law, 
whereas the bulges of early-type galaxies follow a pure power law which is not exhibited by 
their parent population. 

We now wish to establish whether these component fits are consistent with the
Sd-Irr or E \msr{} relations. In particular we wish to test the following hypotheses:
\begin{enumerate}[(a)]
\item ETD and Sd-Irr galaxies are associated
\item ETB and ellipticals are associated
\item LTD and Sd-Irr galaxies are associated
\item LTB and ellipticals are associated
\item LTB and Sd-Irr galaxies are associated
\item LBS and ellipticals are associated
\item LBS and Sd-Irr galaxies are associated
\item S(B)ab-S(B)cd systems fit with a single component (see Sec.~\ref{sec:recon}) and Sd-Irr galaxies are associated 
\end{enumerate}

\begin{figure*}
\centering
\includegraphics[width=0.95\textwidth]{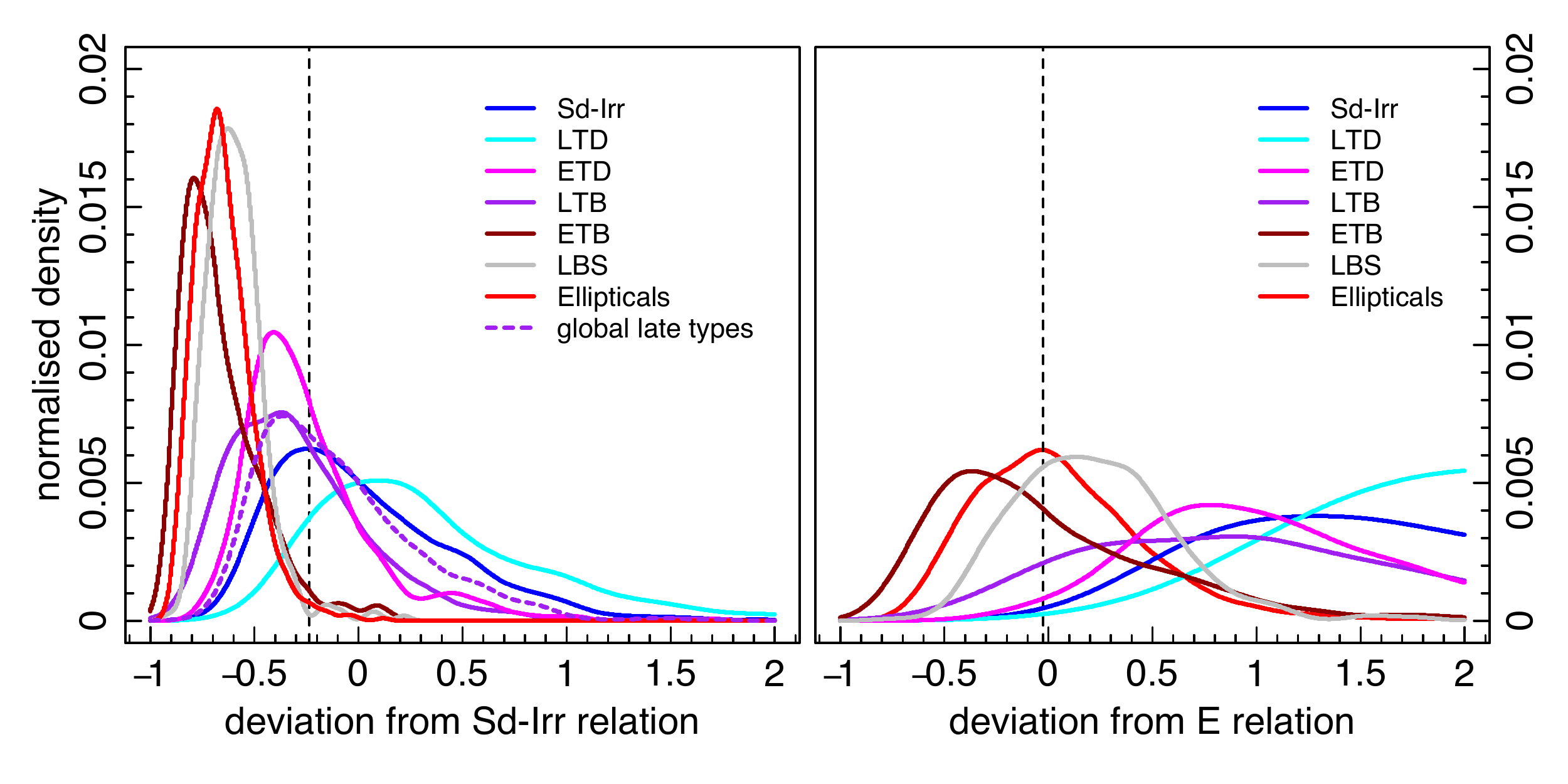}
\caption{Shown are the relative fractional deviations of the different components from the best fit \msr{} relation of the tested parent population.
The left plot shows the deviation of the data from the Sd-Irr \msr{} relation and the right 
plot shows the deviation from the elliptical relation. 
In both plots the area under the curve has been normalised over the range of deviations shown in the plot.}
\label{fig:devs}
\end{figure*} 
 
 In Fig.~\ref{fig:devs} we visualise the affiliation of various components with
 either the Sd-Irr (left) or ellipticals (right)
by plotting their relative deviations, defined as 
\begin{equation}
\label{eq:dev}
 (R_{\rm observed}-R_{\rm predicted})/R_{\rm predicted}
\end{equation}
where $R_{\rm observed}$ represents the sizes of the tested populations,
and $R_{\rm predicted}$ represents the predicted size using either the
Sd-Irr or elliptical \msr{} relation.
In both panels the area under the curve has been normalised and
each population is colour coded as shown in the legend with the dashed, vertical 
black line showing the peak of the deviations for the Sd-Irr galaxies and
ellipticals, i.e. the populations to which we compare.
 The disc components, late-type bulges and the global late-type populations
all broadly align with the Sd-Irr relation. The late-type discs tend to higher deviations
indicating that they are larger than the Sd-Irr population, however, their peak deviation is close to 0. 
The good agreement of the late-type
bulges with the Sd-Irr relation hints at their possible `pseudo'-bulge nature which is corroborated by 
their S\'ersic index distribution. On the other hand dust in latet-type galaxies can artificially lower 
the S\'ersic index of LTB, making them appear larger and thus fall within the parameter space of the
Sd-Irr relation. 
Ellipticals, early-type bulges and LBS on the other hand do not align with the 
Sd-Irr relation and their distributions are completely offset.
Conversely, we find that the LBS and early-type bulges visually align with the elliptical \msr{} relation. They have similar peaks, however, the early-type bulges have a broader distribution, 
which is in good agreement with the larger scatter observed in Fig.~\ref{fig:bd1}. 
Whether this is intrinsic or a byproduct of the decomposition is unclear.
Late-type bulges, discs and Sd-Irr galaxies are offset from the elliptical distribution and do not
follow their relation.\\

 In addition to the qualitative nature of Figure \ref{fig:devs} we also
perform a two sample Kolmogorov-Smirnoff-test (KS-test) for each hypothesis. 
For this we compare the different samples to the \msr{} relation of either Sd-Irr or
elliptical galaxies. 
Since the KS-test in essence compares the cumulative 
distributions of two populations in one dimension it does not take into account the spread of our data around the \msr\ relation. Hence we decided to bin our data in $\Delta\log(M_{\star})$=0.2 steps to  establish the median mass and size of the bin and we use the median bin mass to calculate the 
expected size based on either the Sd-Irr or elliptical \msr{} relation. This
also allows us to test whether the expected size distribution from the \msr\ relation fit agrees with the observed median size distribution of the sample, even for cases which have only little or no overlap in the
mass-size plane (e.g. LBS galaxies and the elliptical \msr\ relation).
The resulting KS statistics are shown in Table \ref{tab:kstest}. 
For our tested assumptions, combining the results of Fig.~\ref{fig:devs} and the KS-test,  we conclude:
\begin{itemize}
\item a,b,c,e,f,h = True
\item d,g = False
\end{itemize}

\begin{table}
\centering
\begin{tabular}{|l|l|l|l}\\
	\hline
Case 			&&	D 		& p-values   \\ 
\hline \hline
\\
\multicolumn{4}{l}{components vs Sd-Irr and E relation}\\
	\hline 	
a) ETD vs Sd-Irr 		&& 0.4 & 0.418\\
b) ETB vs E  			&& 0.38 & 0.66\\
c) LTD vs Sd-Irr   		&& 0.29 & 0.635\\
d) LTB vs E    			&& 0.5 & 0.1\\
e) LTB vs Sd-Irr     	&& 0.42 & 0.256\\
f) LBS vs E	   			&& 0.43 & 0.575\\
g) LBS vs Sd-Irr	   		&& 1 & 0.001\\
\hline
\\
\multicolumn{4}{l}{global fits vs Sd-Irr and E relation}   \\ 
\hline 
Sd-Irr vs Sd-Irr	   				&& 0.33 & 0.73\\
global late-types vs Sd-Irr		&& 0.23 & 0.898\\
global early-types vs Sd-Irr  	&& 0.57 & 0.019\\
\\
E vs E	 					&& 0.2 & 0.994\\
global late-types vs E	   	&& 0.62 & 0.013\\
global early-types vs E   	&& 0.29 & 0.635\\
\hline
\end{tabular}
\caption{Shown are the D- and p-values for a two tailed KS-test on the 
hypothesis stated in Sec.~\ref{sec:strucMSR}.}
\label{tab:kstest}
\end{table}

\subsection{Should S(B)ab-S(B)cd systems be described as single or multi-component?}
\label{sec:recon}

In the previous section we found that both late-type discs and 
late-type bulges are associated with the Sd-Irr relation.
 Additionally, the S\'ersic index distribution of these components 
 (Fig.~\ref{fig:colourn}, bulge, top and disc, bottom) show 
 that late-type discs extend to lower S\'ersic indices than typically 
 expected. In fact, the LTD in our sample have a median S\'ersic
 index of $n \sim 0.6$ (i.e., a more Gaussian-like light profile). 
 The S\'ersic index distribution of our late-type bulges 
 peaks at $n \sim 2$ which is much lower than would be expected for an 
 intermediate to high-mass classical bulge where n$\sim$4. 
 There are two possible reasons, (i) we are seeing the effects of dust affecting both bulge and disc,
 and/or (ii) late-type systems are composed of pseudo-bulges and discs.
 
 Seeing a bulge or disc through dust has the effect of lowering the
 S\'ersic index as well as making them appear larger \citep{Pastrav2013a,Pastrav2013b}.
  On the other hand, if the late-type galaxies do indeed contain pseudo-bulges,
  which are arguably perturbations of the disc, 
 this begs the question as to whether the late-type galaxies should or 
 should not be decomposed into 2 components.

Figure \ref{fig:devs} (left panel) and a KS-test show that the global 
(i.e. single S\'ersic fit) late-type \msr{} relation can 
also be associated with the Sd-Irr relation and a decomposition of the 
late-type 2-component systems is not strictly necessary. 
Another issue to consider here is that if S(B)ab-S(B)cd galaxies are truly
2-component systems than fitting their light profile with a single component only
would bias our size estimation to larger sizes \citep{Bernardi2014}, especially for brighter
and larger galaxies.
However, the majority of our S(B)ab-S(B)cd systems are comparatively small when considering this 
effect found by \cite{Bernardi2014}. Furthermore fitting a final disc relation using 
either global or component fits for the S(B)ab-S(B)cd galaxies 
we find that the resulting \msr{} relations are nearly 
identical (these relations are also given in Table \ref{table:compfits} for reference). 
Hence, as considering either single or two-component sizes has little effect on the \msr{} relation 
and as we cannot conclusively tell the difference between a pseudo-bulge 
and a classical bulge without kinematic data, we opt to use the global (single component)
S(B)ab-S(B)cd galaxies for our final $z=0$ disc relation to avoid over-interpreting our results.
This is also in concordance with other recent studies where late-type two component 
galaxies are considered `bulgeless' discs \citep[for example][]{Sachdeva2015}.

\subsection{Combined disc and spheroid \msr{} relations}
\label{sec:ELT}
\begin{figure}
\centering
\includegraphics[width=0.46\textwidth]{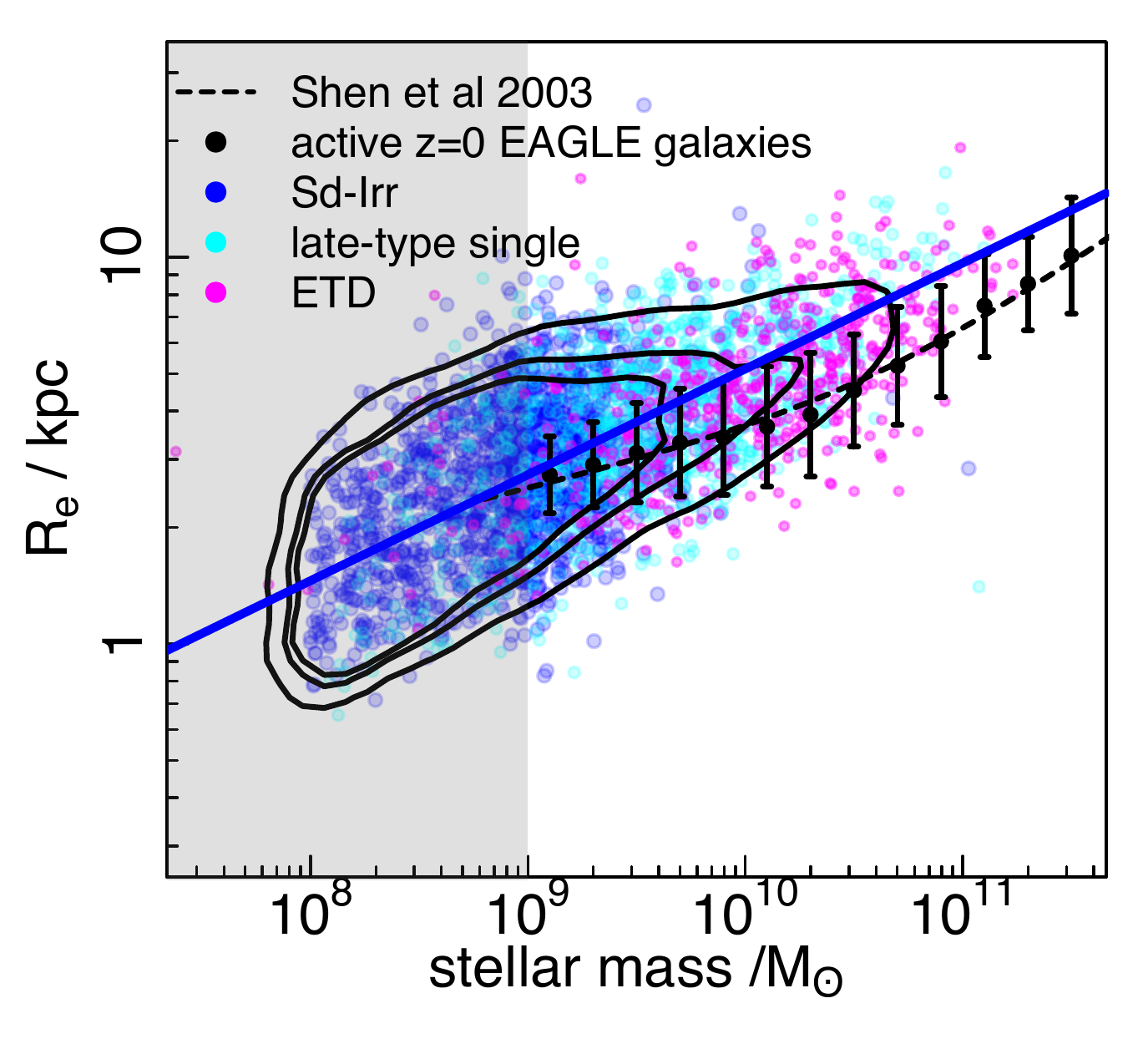}

\includegraphics[width=0.46\textwidth]{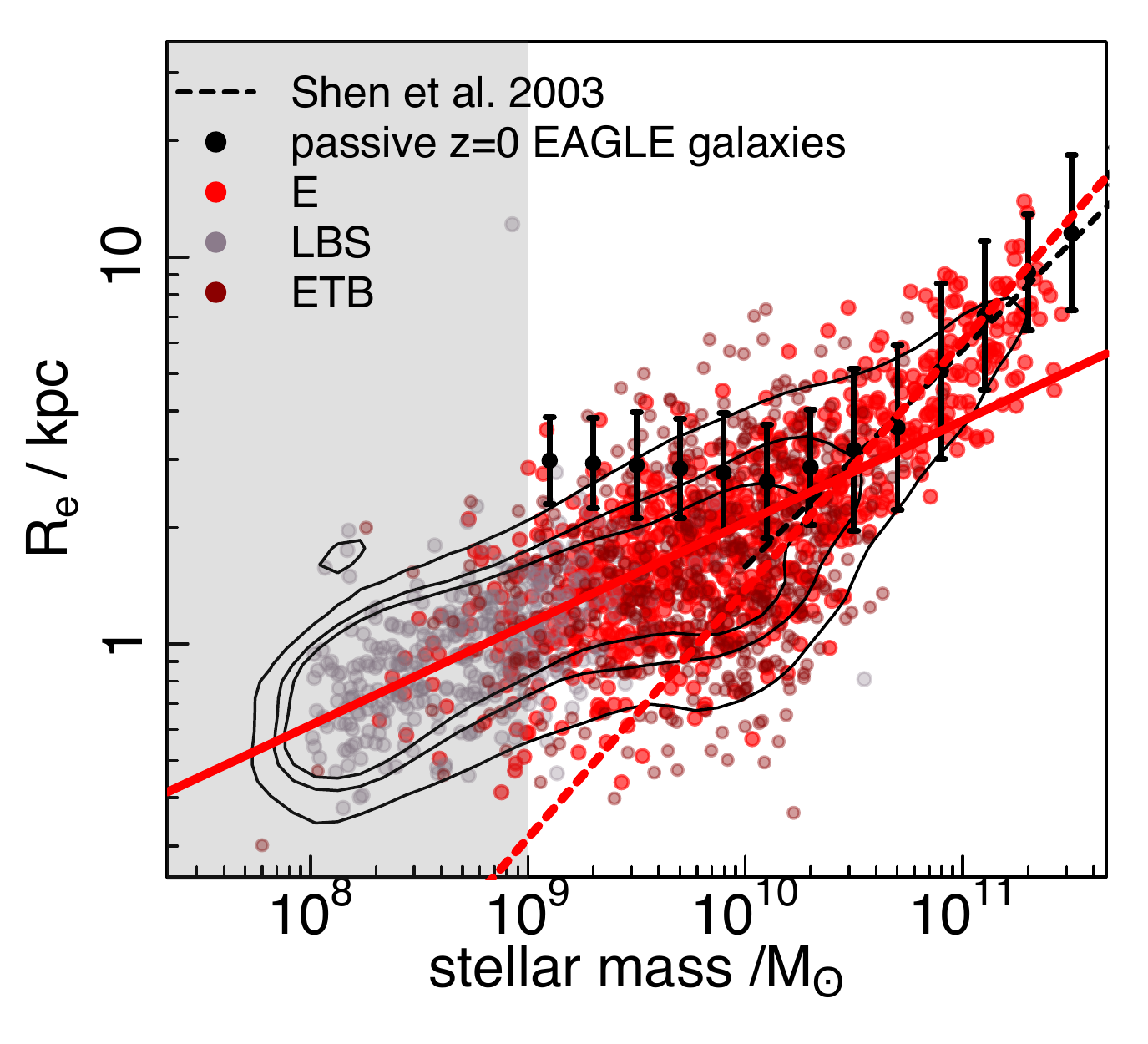}
\caption{ Shown are the final $z=0$ disc (top) and spheroid (bottom) \msr{} relations. 
The disc relation consists of Sd-Irr galaxies, early-type disc components and 
reconstituted late-type 2-component systems. The spheroid relation is composed
of ellipticals, early-type bulges and LBS.
The solid coloured lines are the final disc and spheroid \msr{} relation fits to the data, 
the dashed red line shows the high mass elliptical only () relation,
the grey shaded areas indicate where our smooth 
volume limited sample selection starts and galaxies are up-weighted by their V/V$_{max}$. 
The black (solid) lines are the 90th, 68th and 50th percentiles of the respective mass--size distributions.
Additionally we show the active (top) and passive (bottom) $z=0$ galaxies from the \eagle\ simulation 
(black points + error bars) and the \msr{} relation from \citet{Shen2003} corrected for waveband and 
circularised radius (dashed black line, only shown over the mass range in which it was established).}
\label{fig:bd3}
\end{figure}
In this section we now aim to establish the definitive $z=0$ disc and spheroid \msr{} 
relation composed of associated global and component populations as identified 
in the previous section. In summary, we consider the following populations to be associated:\\
i) Sd-Irr and late-type galaxies and the discs of early-type galaxies and,\\
ii) Ellipticals, early-type bulges and LBS.\\

We show our final combined \msr{} relations for discs (top) and spheroids (bottom) in Figure \ref{fig:bd3}. The data points are colour coded by the population they belong to.
The solid lines show our final disc and spheroid \msr{} relation fits. The dashed red line in the lower panel shows our high mass elliptical relation, which would be more appropriate to use for comparisons with high redshift data (see Sec.~\ref{sec:highz}). In comparison we also show 
simulated $z=0$ \eagle\ galaxies which are actively star-forming (top) and passive (bottom) as black points with error bars. We discuss the selection and comparison of the \eagle\ data in the next Section.
Additionally, we show the \msr{} relation for late- and early-type galaxies by \cite{Shen2003} as a dashed black line. We show these relations as they were used to calibrate the simulated data from \eagle.
Note, we plot the \cite{Shen2003} relations only over the mass range for which they were established and
we have corrected them for the size-wavelength dependence using the equations given in L15 (their Table 4), 
assuming $\mathcal{M}_\star=10^{10}\mathcal{M}_\odot$. Additionally, 
we also correct for the fact that \cite{Shen2003} use a circularised radius\footnote{$R_{c} = \sqrt{b/a} R_{e}$, where b and a are the semi-minor and semi-major axes. Using the average ellipticity of our sample we get $R_e = 1.33 \times R_{c}$.}.
The \msr{} relation parameters can be found in Table \ref{table:compfits} (iv).

For the combined disc populations (Fig.~\ref{fig:bd3}, top) we find that the relation flattens 
considerably from the Sd-Irr only relation, an effect of including the high-mass early-type 
discs and late-type galaxies.
 If our assumption is true that the Sd-Irr galaxies, early-type discs, 
 and `reconstituted' late-type galaxies are related than this could hint at a 
 possible change in the slope of the late-type \msr{} at high masses.
 Comparing our \msr{} relation to the \cite{Shen2003} relation it is obvious that our data
follows an opposite trend at higher masses and does not turn up but down. This effect arises because  
 we compare a component relation to a global relation, we see a similar steepening of our data if only
 global sizes are considered in the \msr{} distribution instead. In addition, since our data
 extends to lower masses than the \cite{Shen2003} analysis we can also see that the
 \msr{} relation does not actually flatten out at lower masses. Overall this shows that
 a linear fit (in log-log space) to the data is sufficient to describe the disc \msr{}
 relation over the observed stellar mass range.

Compared to the elliptical only relation, the slope of the combined elliptical, LBS and ETB \msr{} relation 
also flattens and essentially lies parallel to the final disc relation. 
We also see a turn off in the data, albeit in the opposite sense to the combined disc relation, i.e. the data flattens at the low-mass end.
 For the spheroid \msr{} relation this flattening is caused by the inclusion of more 
low-mass components (mainly LBS), this is corroborated by a comparison of our data to the 
\cite{Shen2003} relation. We see a very good agreement over the
mass range in which the \cite{Shen2003} \msr{} relation was established. However, a clear turn-off,
or flattening can be seen at low masses.
That the flattening is caused  by low-mass galaxies is further supported by comparing 
the final spheroid relation to the low-mass elliptical only relation, see Table \ref{table:compfits} 
(ii), which are within the errors identical. 
 Additionally we fit the \msr{} relation to ellipticals and early-type bulges only and find 
 that this relation is, within the errors, identical to the elliptical only relation, 
 supporting the notion that early-type bulges are indeed classical bulges and ``elliptical-like''.
 This begs the question whether the LBS are all indeed early-type galaxies.
As shown in Fig.~\ref{fig:devs} the distribution of the relative deviations of LBS from the 
elliptical \msr{} relation is consistent with them being associated. Their broad
distribution, however, which looks similar to a top-hat function, extends to high deviations which
 suggests that not all LBS galaxies are the same. 
Without higher resolution imaging data we cannot yet conclusively say if LBS are indeed all 
early-type galaxies. Preliminary visual inspection of LBSs available in the higher resolution 
VIKING imaging shows that a significant fraction of them are actually two-component systems.

If the LBS and ETBs, as well as the high- and low-mass ellipticals are all 
indeed the same population, then a curved fit, i.e.~a double power law, is necessary to 
describe the \msr{} relation. 
In lieu of a definitive answer we recommend considering the high- and low-mass early-type 
populations separately. In Table \ref{table:compfits} (ii) we provide a \msr{} relation 
for high- and low mass ellipticals considering two different mass separators at 
$M_{\star}=10^{10}M_{\odot}$ and $M_{\star}=2 \times 10^{10}M_{\odot}$. The combined 
\msr{} relation to all ellipticals, ETBs and LBS in Table \ref{table:compfits} (iv) should 
only be considered for samples that primarily contain galaxies with 
$M_{\star} \leq 10^{10}M_{\odot}$. Furthermore, it should be noted that previous studies have 
reported a second deviation of the early-type \msr{} relation from a simple power law at very high masses
\citep[$\mathcal{M}_{\star} \sim 2\times 10^{11} \mathcal{M}_\odot$, e.g.,][]{Bernardi2007,Hyde2009,Bernardi2011,Bernardi2014}.
Due to the limited volume we survey, we do not see this deviation. However, as this second change in slope is likely linked to
the formation history of galaxies it needs to be taken into account when studying the high-mass end of the \msr{} relation. 

Finally we include the caveat that the flattening 
of both the disc and spheroid population could be real or due to miss-classification in our 
visual morphology sample or a significant bias in our \bd{} decomposition. 
Ultimately deeper data, such as that provided by VST KiDS and a fit in all available wavelengths
 to establish robust masses should clarify whether the
flattening is real or an artifact of our methodology.

\section{Does the size distribution pose a problem for simulations?}
\label{sec:sims}

\begin{figure*}
\centering
\includegraphics[width=0.95\textwidth]{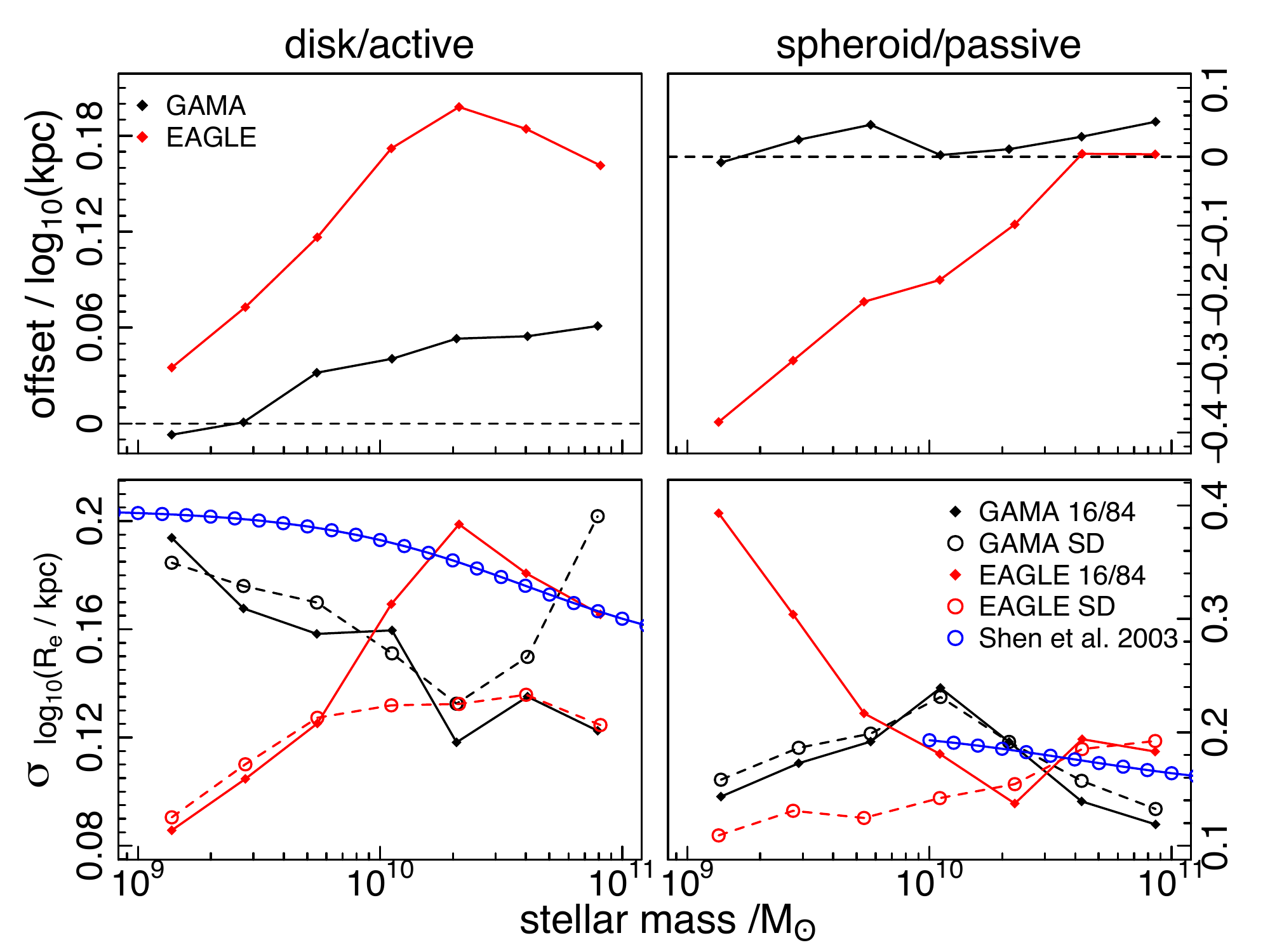}
\caption{Shown is the offset (top) of the Gaussian distribution fit of the observational 
and simulated data from the disc (left) and spheroid (right) \msr{} relation.
The bottom panel shows the corresponding scatter $\sigma_{\log(R_e)}$ (SD, circle plus dashed lines) 
as well as the 16$^{th}$/84$^{th}$ percentile scatter (16/84, filled symbols plus solid line) 
of the distribution. Additionally we show the $\sigma$--mass relation from \citealt{Shen2003} for comparison.}
\label{fig:simcomp}
\end{figure*}

In this section we have a first look at comparing our final disc and spheroid relations with data 
from the \eagle\ simulation \citep{Schaye2015,Crain2015}. 
\eagle\ is a suite of cosmological hydrodynamical simulations 
performed at two numerical resolutions,
in periodic volumes with a range of sizes, and using a variety of 
subgrid implementations to model physical processes below the resolution limit. 
The subgrid parameters governing energetic feedback mechanisms of the \eagle\ reference model were
 calibrated to the $z=0$
galaxy stellar mass function, galaxy stellar mass - black hole mass relation, 
and galaxy stellar mass - size relations (see \citealt{Crain2015} for details and motivation).
 The \eagle\ reference model reproduces many observed galaxy
relations that were not part of the calibration set, such as the evolution of the galaxy stellar
mass function \citep{Furlong2015}, of galaxy sizes (\citealt{Furlong15b}), of their
optical colours \citep{Trayford15}, and of their atomic (\citealt{Bahe16}) and
molecular gas content \citep{Lagos15}, among others, and thus is an excellent testbed to 
compare with our observations.

We use the public database of \eagle\ described in \citet{McAlpine2015}. 
In particular, we focus our attention on the reference model of \eagle\ run in
a cubic volume of length $100$~ comoving Mpc on a side with $2\times 1504^3$ dark 
matter and gas particles (particle masses are $9.7\times 10^6 \mathcal{M}_\odot$ 
and $1.81\times 10^6 \mathcal{M}_\odot$, 
respectively, which help to reach a physical 
resolution of $0.7$~kpc. One of the notable aspects of \eagle\ is the plethora 
of sub-grid baryonic physics included in the 
model: (i) radiative cooling and photo-heating rates, (ii) star formation, 
(iii) stellar evolution and metal enrichment, 
(iv) stellar feedback, and (v) black hole growth and AGN feedback. 
These physical models are the key ingredient to reproduce 
a large set of properties of the observed galaxy population in the local Universe. 
For more details of the simulation we refer the reader to \citet{Schaye2015}.   \\

To compare \eagle\ data with our observations we use the masses and sizes 
of z=0 active and passive galaxies, calculated from data in the
public database, following the method described in \cite{Furlong2015}. 
Briefly, the size is the mean half-mass radius taken within a 100kpc aperture
projected along the x-y, x-z, and y-z axes. Note that this definition of size differs
from the size used to calibrate the simulation against the \cite{Shen2003} late type galaxies.
The galaxy sizes used in the calibration are based on the S\'ersic scale length
obtained from fitting S\'ersic profiles to the surface density profiles of the simulated galaxies.
To separate active and passive galaxies  \cite{Furlong2015,Furlong15b} use a specific 
star-formation rate cut of 0.01 $Gyr^{-1}$, which is approximately one decade below the 
observed main sequence of star formation.
For the following comparison there are two notable caveats:\\
(i) we compare our measurements of the disc and spheroid \msr{} relation for the
half-light radius to the half-mass radius of the simulated galaxies, and\\
(ii) we separate active and passive simulated galaxies to compare to our discs and bulges, respectively. 
We caution the reader that although a correlation between being bulge (disc)-dominated 
and being passive (active) is expected, they are not necessarily the same populations.\\

Figure \ref{fig:bd3} shows our final disc and spheroidal \msr{} relations.
We plot the \eagle\ data as black points and the error bars indicate 
the 16$^{th}$ and 84$^{th}$ percentile of
their distribution. We also show the \cite{Shen2003} relation 
for $n<2.5$ and $n>2.5$ galaxies as the dashed line.
The difference in the shape of the disc \msr{} relation (Fig.~\ref{fig:bd3}, top) is in part caused 
by comparing disc only components (our data) with global sizes (\eagle). 
This is because also considering a bulge component in a global profile fit has the effect 
of lowering the half-mass radius due to their typically smaller size and higher 
concentration compared to discs.
For the spheroid relation (Fig.~\ref{fig:bd3}, bottom) we agree well with the \eagle\ 
data down to $\mathcal{M}_*\sim 10^{10}\mathcal{M}_{\odot}$.
Below this mass limit we see a  much less marked change of slope compared to the \eagle\ data.
This could be due to comparing passive galaxies in \eagle\ with our spheroid sample that 
contains LBS galaxies. The latter are arguably active systems. 
On the other hand, at least in part, this could be explained by the known limitations of the 
simulation, e.g. the high fraction of passive low mass galaxies at $z=0$ due to the finite 
sampling of the star-formation in low mass galaxies.  
A one-to-one comparison,
in which \eagle\ galaxies are analysed with the same pipeline applied to GAMA galaxies is needed
to shed light on this issue. A further point to consider is the shape of the \msr{} relation 
at very high masses ($\mathcal{M}_*\geq 10^{11.3}\mathcal{M}_{\odot}$) which has been shown to deviate
from a pure power law \citep[][e.g.,]{Bernardi2007,Hyde2009,Bernardi2011,Bernardi2014}. As mentioned
previously, due to our small survey volume, we do not sample the high-mass end of the \msr{} 
relation well and cannot confirm the curvature of the relation. However, to ensure that simulations
return realistic galaxy sizes at all masses, not only the mass range over which they were matched, a 
detailed comparison with the relevant studies is necessary.

Figure \ref{fig:simcomp} summarises the findings of Figure \ref{fig:bd3}.
Here we compare the distribution of the vertical scatter for the GAMA and \eagle\ data from our
disc (left) and spheroid (right) \msr{} relation.  We divide our sample in bins
 of 0.3 dex in stellar mass and fit a Gaussian to the distributions (see Figures 
 \ref{fig:logrescatter} and \ref{fig:logrescatter2}) to study the scatter of the 
 data from the \msr{} relation at fixed stellar mass. 
 The top panel of Fig.~\ref{fig:simcomp} shows the offset ($\mu$ of the Gaussian fit) 
 of the $\log (R_e)$ scatter from the \msr{} relation. 
 Note that for the high mass spheroid/passive galaxy sample
 we actually compare to the high mass ($\mathcal{M}_*>10^{10}\mathcal{M}_{\odot}$) elliptical 
 \msr{} relation. This does not change the scatter of the data but has the effect of moving 
 the offset from the \msr{} relation to $\sim 0$.
 The bottom panel shows the $\sigma_{\log(R_e)}$ versus stellar mass distribution.
 We show the sigma (standard deviation, SD) of the best fit Gaussian to the underlying distribution as well as the
 16$^{th}$ and 84$^{th}$ percentile scatter of the GAMA and \eagle\ data. In cases where the
 underlying distributions are somewhat skewed the SD and 16/84$^{th}$ percentile sigma 
 do not agree well and could even hint at a possible bimodality in the underlying
 distribution. We also include the scatter versus mass relation from \cite{Shen2003}, corrected 
 from $\log_e$ to $\log_{10}$, for comparison. Note that the \cite{Shen2003} scatter--mass relation
 is based on the combined scatter of their early- and late-type \msr{} relations.
 In Appendix \ref{app:sim_comp} we show the histograms of the scatter of the data from 
 the \msr{} relation for each mass bin with the best fit  distribution overplotted.
  
Examining Fig.~\ref{fig:simcomp} the offset in the modes of the GAMA and \eagle\ data is expected, 
as seen in Figure \ref{fig:bd3}. This is due to the
\eagle\ simulation being calibrated using the \cite{Shen2003} relations.
To some extent the variance (or sigma values as indicated on the Figure) is of more interest 
as these have not been explicitly tailored in the simulation to match the data distributions. 

We focus our analysis on the sigma derived from the Gaussian fits, i.e., the points labelled 
SD in Fig.~\ref{fig:simcomp}. We find that for the disc/active population the variance 
in the simulated data is almost always smaller than the observations. 
For the spheroid/passive populations there is a divide at 
$\mathcal{M}_* \sim 2 \times 10^{10}\mathcal{M}_{\odot}$ with the simulated 
data having a smaller variance for less massive systems compared to the data
and a comparable variance for more massive systems.
This is somewhat surprising as the dark matter spin distributions are known to be quite broad. 
If coupling is strong one would expect the distribution of
specific angular momentum and disc sizes to be comparably broad. 
Because we are comparing light against mass and components 
against classes we should be careful with interpreting any deviations. 
Clearly an improved comparison can be made from \bd{} decompositions of the \eagle\ images which 
would place the observational and simulated data onto the same footing.
Although \textit{gri} images of \eagle\ galaxies with $\mathcal{M}_* > 10^{10}\mathcal{M}_{\odot}$ 
have been made publicly available from their database, it is still not sufficient to perform an
analysis like the one done here for GAMA. Images of individual bands, 
preferably to lower stellar masses, would be required for this.
For the moment we consider Figures \ref{fig:bd3} and \ref{fig:simcomp} 
to provide a good demonstration of the potential of the mass-size plane for comparing 
observational and simulated data.

\begin{figure*}
\centering \includegraphics[height=0.9\textheight]{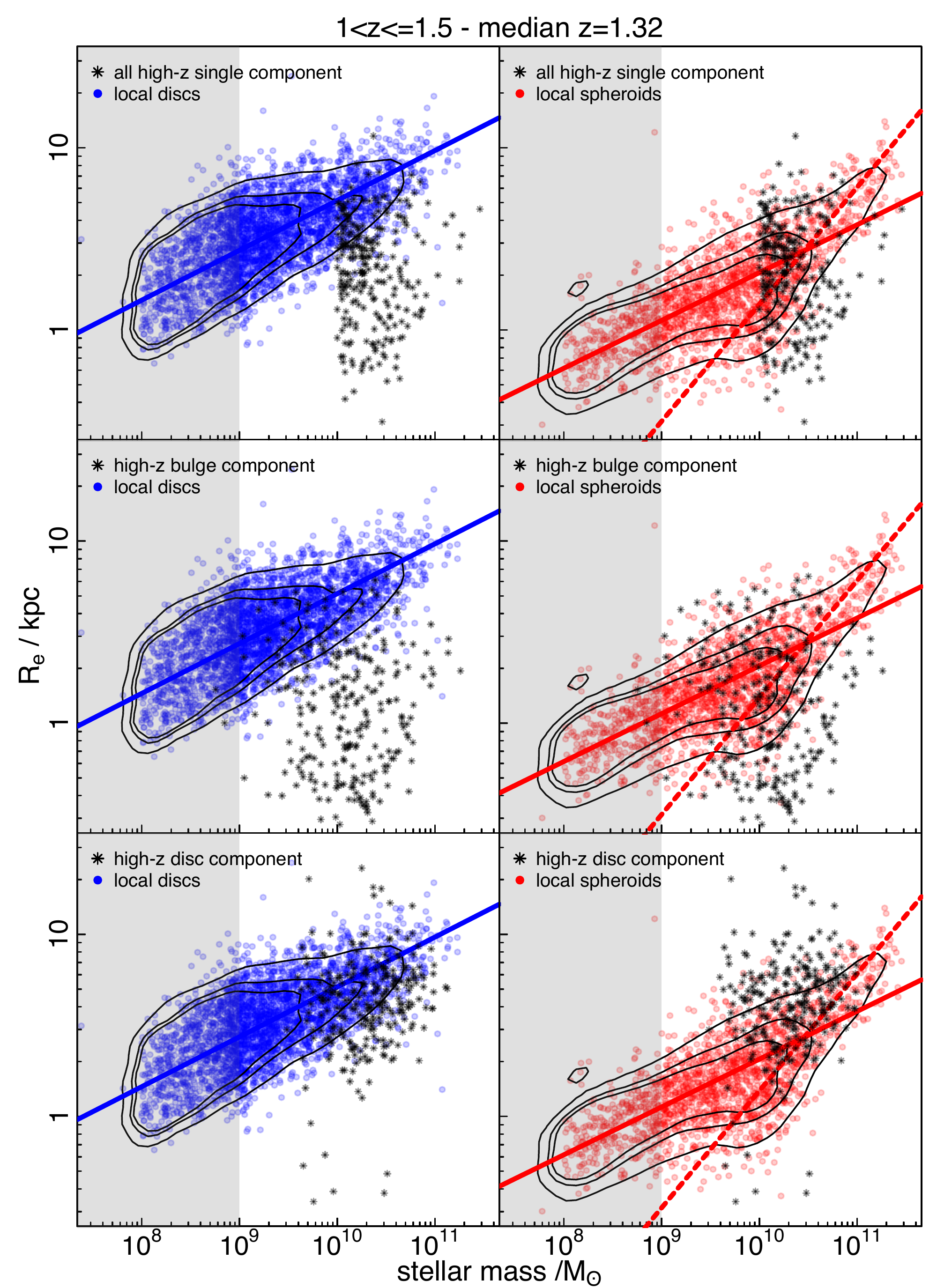}
\captionof{figure}{We show the local disc (blue, left side) and
  spheroid (red, right side) \msr{} distributions in comparison to the
  high redshift CANDELS data ($1<z<1.5$) for our three high-redshift
  populations. The solid blue and red lines are our final disc and
  spheroid \msr{} relation as presented in Sec.~\ref{sec:ELT}. The dashed
  red line is the local high mass ($\mathcal{M}_*>10^{10}\mathcal{M}_{\odot}$) 
  elliptical \msr{} relation as given in Table \ref{table:compfits} (ii).
  It is immediately obvious that most data do not agree
  with the disc \msr{} relation. On the other hand there is good
  agreement between our local spheroids and the high-redshift single 
  component systems and the high-redshift bulge components 
  (with a compact extension of the high-redshift
  bulge components).}
\label{fig:highz}
\end{figure*}

\section{Comparison with high redshift data}
\label{sec:highz}

\noindent There is a well know discrepancy between the \msr{}
relation of high redshift galaxies (z$>$1) and the local \msr{}
relation, with the high redshift galaxies at the same mass being
smaller than their low redshift counterparts \citep[e.g.~][]{Daddi2005, Longhetti2007, vanDokkum2008,Trujillo2006, Trujillo2007}. This itself is not a problem
since an evolution in the \msr{} relation is expected due to 
galaxies at later times being formed through less dissipative
events (i.e. low redshift progenitors are gas-poorer than high redshift progenitors,
see e.g.~\citealt{Hopkins2009}). Hence at lower redshift newly formed galaxies are expected to
be bigger than their high redshift counterparts. However, the lack of old
massive compact systems at low redshift means that the galaxies
observed at high redshift must have grown by a factor of up to
$\sim$5-6 to end up on the local \msr{} relation.
One might argue that the measurements of high redshift galaxies are inherently difficult and
the observed size growth is biased by systematics. However, several studies have shown that, even
considering all the uncertainties in the mass and size measurements of the high-z galaxies,
the size growth is real. Even for the most unfavorable cases the high redshift galaxies lie well below the present day \msr{} relation \citep[see e.g.][]{Buitrago2013,Weinzirl2011}.\\

In this section we briefly compare published measurements of
high-redshift galaxies from the CANDELS-UDS field
\citep{Mortlock2013,Mortlock2015,MB2015}, against our local disc and
spheroid \msr{} relations. To briefly summarise, the CANDLES-UDS data
contains 1132 galaxies with $\mathcal{M}_*>10^{10}\mathcal{M}_{\odot}$
and 1$<$z$<$3.  Of these 683 are fit with a single component S\'ersic
profile while 449 with a \bd profile.  For the two-component fits the
disc is set to $n=1$ and the bulge $n$ is free unless the fit failed
in which case the bulge was reset to either $n=1$ or $n=4$
\citep{MB2015}. 

The CANDELS-UDS data were obtained using the Hubble
Space Telescope's Wide Field Camera 3 (WFC3\slash{}IR) F160W ($H$)
band, which is well matched in physical resolution and rest-wavelength
to the GAMA low-redshift SDSS data. Here we restrict ourselves to the
redshift range $1 < z < 1.5$ where the $H$-band equates to a
comparable rest wavelength of ~$640-800$nm. For the HST data stellar
masses were derived by the CANDELS team using BC03 stellar
populations, a Chabrier IMF and with the same $\Lambda$CDM cosmology
as our analysis. One minor difference worth highlighting between the
HST and GAMA analysis, is the derivation of component masses. For the
HST data this was based on the single band bulge-to-total ratio rather
than component colours (see Section \ref{sec:compmass}).\\

The high-redshift CANDELS-UDS sample naturally divides into three distinct
populations: single component systems, bulges of two-component
systems, and discs of two-component systems.  We now explore whether
any of these three populations follow a similar \msr{} relation to the
local benchmarks.\\

\begin{figure*}
\centering
\includegraphics[width=0.95\textwidth]{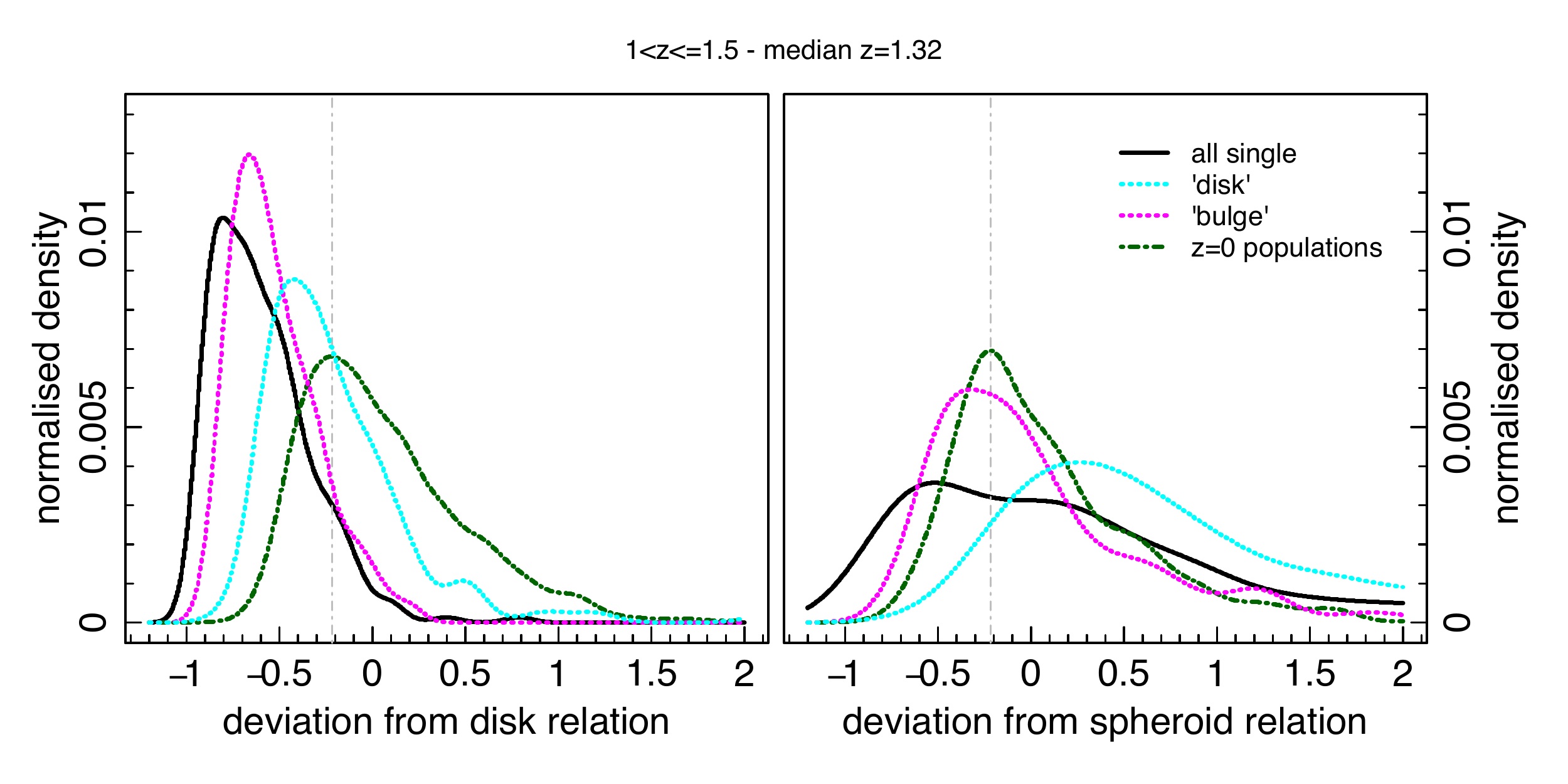}
\captionof{figure}{Shown is the same comparison as Fig.~\ref{fig:devs} but for the CANDELS high redshift data ($1<z\leq1.5$) compared to our final disc (left) and high mass elliptical (right, $\mathcal{M}_*>10^{10}\mathcal{M}_{\odot}$) \msr{} relations.}
\label{fig:devs2}
\end{figure*}
 
In Fig.~\ref{fig:highz} we show the \msr{} plane with
the various high-redshift samples overlaid on the local benchmark data (left
panel against discs and right panel against spheroids). Note that
we show both the local spheroid \msr{} (solid red line) and the high
mass elliptical \msr{} relation ($\mathcal{M}_*>10^{10}\mathcal{M}_{\odot}$,
dashed red line). For our comparison we concentrate on the high mass
elliptical relation since this describes our low redshift data
better in the mass range observed for high redshift galaxies.
It is immediately apparent that none of the populations provide a 
good association with the low-redshift disc benchmark (left panel), 
with all systems being significantly more compact. 
However, the high-redshift discs lie the closest and fall just below the low
redshift relation (bottom left panel). On the other hand, we see what appears to be a fairly 
close association between the high redshift single component systems 
and the low redshift spheroid benchmark (top right panel). 
The high redshift bulges also overlap with the low-redshift-spheroid
relation (middle right panel), but extend to significantly lower sizes
as well.

To elucidate these issues further, we show in Fig.~\ref{fig:devs2} the deviation distribution. 
We follow our methodology from before (see Section \ref{sec:strucMSR}) and use
Eq.~\ref{eq:dev} to calculate the deviation from the local benchmark relations.
Here $R_{\rm observed}$ represents the sizes of the CANDELS-UDS data,
and $R_{\rm predicted}$ represents the predicted size using either the
low redshift disc relation or low redshift high mass ($\mathcal{M}_*>10^{10}\mathcal{M}_{\odot}$) 
elliptical relation \msr{} relation. 
On Fig.~\ref{fig:devs2} each of the three populations are shown compared 
against either the local disc (left panel) or local spheroid (right panel) relations.
The dark green dot-dashed lines show the scatter of the local data
about the disc and spheroid relations. Note that the low redshift spheroid
deviation (dark green dashed line, right panel) is only evaluated down to
$\mathcal{M}_*=10^{10}\mathcal{M}_{\odot}$ to give a fair comparison, since
a high mass only \msr{} relation is used for evaluation. The solid black line shows the
high-redshift single component sample while the dotted lines show the high-redshift
discs (cyan) and bulges (magenta). Fig.~\ref{fig:devs2} reiterates
our findings from Fig.~\ref{fig:highz}. Compared to the
local spheroid benchmark, two populations show a plausible fit, the
single component systems, and the high redshift bulges,
albeit with a greater spread potentially indicative of the greater
measurement error associated with fitting high redshift data. Whereas
none of the distributions agree with the disc \msr{} relation.
However, the high redshift disc components have the largest overlap
with our local disc \msr{} relation.

In addition to the qualitative nature of Figures \ref{fig:highz} and
\ref{fig:devs2}, we also perform a KS-test, as described in
Sec.~\ref{sec:2comp}, to establish the association of the high
redshift data with our local disc and high mass spheroid \msr{} relation.  
The resulting test statistics are shown in Table \ref{tab:kstest2}, and
corroborate our visual inspections.\\

Putting aside external observations (e.g. visual morphology used as priors), 
our analysis suggests that the majority of high-redshift systems overlap 
satisfactorily with the low-redshift spheroid \msr\ relation, with the exception 
of the high-redshift disc components.  
The obvious and simplest conclusion, is that we are
essentially seeing bulge and spheroid formation/emergence at high-redshift,
with some two-component systems existing, which adhere reasonable
closely to the $z=0$ spheroid relations but with the high-redshift discs
somewhat more compact than their $z=0$ counterparts. This
observation meshes well with the notion of rapid spheroid formation
at high-redshift \citep[$z>1.5$;][]{Tacchella2015} followed by disc growth at
intermediate to lower-z \citep[$z<1.5$; e.g., ][]{Sachdeva2015}, i.e., two-phase evolution as
described in \cite{Driver2013}. The obvious objections, however,
are that observations of high-redshift systems generally show them
to be visually clumpy, vigorously star-forming, and exhibiting clear
evidence of systemic rotation, and hence are often described as
disc-like 
(see for example \citealt{Forster2006,Forster2009,Green2010}, and the recent 
review by \citealt{Glazebrook2013}). However,
these observations also show strong vertical velocity dispersions
\citep[e.g.~][]{Barro2014}, verging on or exceeding the
Toomre stability criterion \citep{Toomre1964}, i.e.~if these are discs they
are highly unstable and unlike any type of disc seen locally.

At some level there is a semantic issue worth raising: When exactly
does a spheroid become a spheroid, or a disc become a disc? Even in
the ideal scenario of an entirely isolated collapsing gas cloud, it is
likely to go through several star-burst phases, fragmentation and
merging of these fragments before finally resembling what we consider
a classical elliptical. Exactly at what point should we start calling
such a system an elliptical, at the moment of first collapse or only
after all star-formation has ceased and the system becomes dynamically
relaxed? If the critical criteria are along the lines of the Hubble
classification then clearly the high-redshift systems are not spheroids,
however, it is also clear these are not conventional discs (smoothly
rotating systems with minimal velocity dispersions and aspect ratios
of 1:10), and the use of this terminology is equally misleading.

The mass-size relation essentially maps fundamental (conservable)
quantities of mass and angular momentum. In this sense the mass-size
relation is quite powerful and appears to be arguing that the majority
of systems at high-redshift are, if not spheroids, proto-spheroids 
in the process of settling into spheroids. Those systems that
do appear to have two-components exhibit bulges consistent or slightly smaller
than low-redshift bulges (which could be due to increased nuclear activity making the bulges appear more compact), and discs which are offset to lower
sizes, however disc growth is expected to continue to lower redshifts. 

While the above paints a consistent and tantalising picture the
caveats at this stage are significant. The analysis of the CANDLES-UDS
data has been conducted by an independent group using distinct methods
and strategies which could introduce systematic offsets. Distances are
also based on photometric redshifts for the vast majority of the
high-redshift sample. The sample size is also relatively small (subject to
cosmic variance effects), and spans a particularly narrow mass range.
Nevertheless, as we push the depth and area boundaries with facilities
such as Euclid and WFIRST, the \msr{} scaling relation shows great promise for
providing not only a connection to the hydrodynamical simulations but
also as a bridge between the low and high-redshift Universe.

\begin{table}
\centering
\begin{tabular}{|l|l|l|l}\\
	\hline
Case 			&&	D 		& p-values   \\ 
\hline \hline
\\
\multicolumn{4}{l}{(i) high redshift vs local disc \msr{} relation}\\
	\hline 	
a) all high-redshift single 	 &&  1  &  0.002\\
b) disc 					 &&  0.62  &  0.087\\
c) bulge 				 &&  1  &  0\\
\hline
\\
\multicolumn{4}{l}{(ii) high redshift vs local high mass spheroid \msr{} relation}   \\ 
\hline 
a) all high-redshift single 	 &&  0.5  &  0.474\\
b) disc 	 				 &&  0.75  &  0.019\\
c) bulge 				 &&  0.4  &  0.418\\
\hline
\end{tabular}
\caption{Shown are the D- and p-values for a two tailed KS-test on the 
hypothesis that high redshift  $1\leq z\leq 1.5$ galaxies and components 
are associated with either the local disc (i) or spheroid (ii) \msr{} relation.}
\label{tab:kstest2}
\end{table}

\newpage
\section{Summary and Conclusion}
\label{sec:SnC}

 We have presented our bulge-disc decomposition
catalogue for 7506 galaxies from the GAMA survey in the redshift range of 
$0.002<z<0.06$ (Sec.~\ref{sec:2comp}).
To overcome the limitations of the LM minimisation algorithm 
used in \galfit{}, which can get trapped in local minima (especially for 2-component fits), we repeatedly
fit our galaxy sample with varying starting points to map out the parameter space.
For the single component galaxies we use a set of 33 combinations of the starting parameters, and for 
the 2-component fits we use a set of 88 combinations.
We implement a screening process to prune bad fits and determine the final fitting values and errors
from the median of the acceptable fits.
We use the 16$^{th}$ and 84$^{th}$ percentile of the remaining output parameter distribution 
to determine the error on the median model combined with a 10\% error floor.
Through this strategy we reduce our catastrophic failure rate from $\sim 20\%$ to $\sim 5\%$.\\

We then presented the \msr{} relations of our sample by Hubble type and component with the 
component masses based on an estimation from the bulge and disc colours.
Next we explored the association of the bulge and disc components with either the Sd-Irr or 
elliptical \msr{} relation. 
We find that S(B)ab-S(B)cd galaxies likely consist
of a disc plus a pseudo-bulge. Considering that a pseudo-bulge is a perturbation of the
disc we decide that our late-type 2-component systems are best represented by a single 
component S\'ersic fit. Thus we associate elliptical, early-type bulges and LBS for the
spheroid \msr{} relation and Sd-Irr, single component fit S(B)ab-S(B)cd galaxies and early-type discs
for the final disc \msr{} relation, which we provide as a definitive 
low redshift benchmark:\\
\\
$R_e=5.141 \mathrm{\left(\frac{\mathcal{M}_*} {10^{10}\mathcal{M}_{\odot}}\right)}^{0.274}$ for discs and,\\
\\
$R_e=2.063 \mathrm{\left(\frac{\mathcal{M}_*} {10^{10}\mathcal{M}_{\odot}}\right)}^{0.263}$ for spheroids.\\

However, we caution the reader that the spheroid relation is heavily dominated by low mass
galaxies. If a comparison to high mass spheroids is needed then the
high mass elliptical \msr{} relation (see Table \ref{table:compfits}) 
should be used in lieu of a curved spheroid relation.\\

Next we used our local disc and spheroid \msr{} distributions 
to compare to data from the \eagle\ simulation.
We find a qualitatively good agreement between the sizes of the \eagle\ data and our
\msr{} relations.
This is not surprising as the sizes in \eagle\ were calibrated
using the \citep{Shen2003} \msr{} distribution.
Hence the variance is of more interest as this has not been explicitly matched between
the observed and simulated data.
Comparing the scatter of the observed and simulated data we find that in almost all
cases the simulated data has a smaller scatter which is unexpected, 
considering that the dark matter spin distribution is known to be fairly broad and
we would expect the sizes and angular momentum distributions to have similarly broad distributions.
Since we are comparing half-light sizes to half-mass
sizes and components versus active and passive galaxies, we caution the reader 
to not over-interpret this comparisons. Instead we would like to highlight the 
potential of using the mass-size plane to compare observational and simulated data.

Finally, we compare our local \msr{} relations to high 
redshift data from the CANDELS-UDS field. We concentrate on available data
 in a redshift range of $1<z\leq 1.5$ with available single and 2-component fits \citep{Mortlock2013,MB2015}.
We generally find that low-mass high redshift galaxies agree better with the local 
\msr{} distributions than high-mass high redshift galaxies.
Furthermore, high redshift systems, with the exception of disc components, more closely follow that of
our local spheroid relation. The high redshift discs on the other hand follow the local
disc \msr{} relation, albeit offset to slightly smaller size.
 We interpret this as evidence for spheroid formation
at high redshift and propose that further disc formation and/or 
growth does not occur until later times.

\section{Acknowledgements}
RL would like to acknowledge funding from the International Centre for
Radio Astronomy Research and the University of Western Australia.
SB acknowledges the funding support from the Australian Research Council through a Future Fellowship (FT140101166).
CL is funded by a Discovery Early Career Researcher Award (DE150100618).

GAMA is a joint European-Australasian project based around a
spectroscopic campaign using the Anglo-Australian Telescope.  The GAMA
input catalogue is based on data taken from the Sloan Digital Sky
Survey and the UKIRT Infrared Deep Sky Survey.  Complementary imaging
of the GAMA regions is being obtained by a number of independent
survey programs including GALEX MIS, VST KiDS, VISTA VIKING, WISE,
Herschel-ATLAS, GMRT and ASKAP providing UV to radio coverage.  The
VISTA VIKING data used in this paper is based on observations made
with ESO Telescopes at the La Silla Paranal Observatory under
programme ID 179.A-2004.  GAMA is funded by the STFC (UK), the ARC
(Australia), the AAO, and the participating institutions. The GAMA
website is http://www.gama-survey.org/.

\footnotesize
\bibliographystyle{mn2e}
\setlength{\bibhang}{2.0em}
\setlength{\labelwidth}{0.0em}
\bibliography{component_MSR}
\normalsize

\appendix

 \newpage
 \onecolumn
\section{Flagging}
\label{App:flags}

As described in Section \ref{sec:selection} our sample of 7506 $0.002 < z < 0.06$
galaxies has been classified onto the Hubble type system as described
in \citealt{Moffett2015}. The sample contains 860 E, 826 S(B)0-S(B)a,
1421 S(B)abc, 3531 Sd-Irr systems and 868 Little Blue Spheroids (LBS). 
The E, LBS and Sd-Irr's we consider single component systems, best fit by a single 
S\'ersic profile, while we assume the remainder to be best described by a two-component, double
S\'ersic profile.\\

Following the decision on one or two components we flag the resulting fits for various criteria:
\begin{enumerate}
\item very high or low \bt{} (\bt{} $>$ 0.8 and $<$ 0.1 respectively)
\item disc n $>$ bulge n
\item \bt{} reverses between minimum reduced \x and median model
\item bulge and disc position angle offset
\item the minimum reduced \x solution is an outlier to the median value
\end{enumerate}
Flags (i) through (iv) are only evaluated on the sample of two-component galaxies.
Flag (v) is evaluated for all galaxies (i.e., single and two-component).

For our two-component sample we have a total of 962 galaxies with at least one flag ($\sim44\%$). 
This drops drastically when checking for galaxies with several flags, and we find
only 206 galaxies with more than one flag raised.
Our single component sample has a total of 164 flagged galaxies ($\sim 3\%$).

To check whether any of the flags are more likely to produce unsatisfactory fits 
we visually inspect a random sample of 50 flagged two-component galaxies. We found 
that in many cases the median fit is acceptable and only flags (i) and (ii) are 
more likely to yield potentially bad fits.\\

\noindent \textbf{(i) galaxies with high (low) \bt{}}\\
We selected galaxies with a median \bt{} $>$ 0.6 and visually inspected their minimum 
reduced \x{} and median fit results as well as the convergence plots. 
For each galaxy we decided whether it is better fit with a double component or 
single component or if it has a bad or uncertain fit (i.e., the fit has bad apertures 
and no solution can be found, or it is unclear whether a 2-component fit is appropriate).

\begin{figure}
\centering
\begin{minipage}[b]{0.45\textwidth}
   \includegraphics[width=\textwidth]{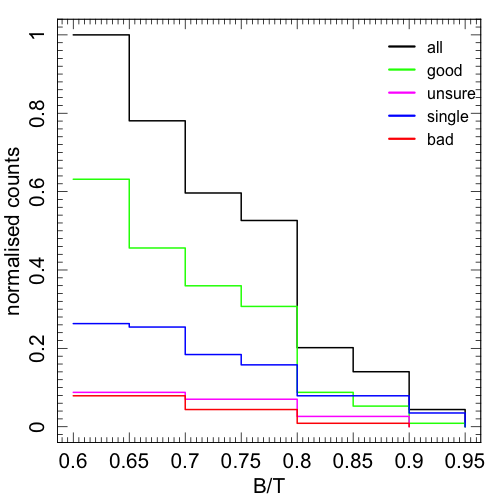}
    \caption{The distribution of galaxies with a median \bt{} $>$ 0.6 and the visual  decision whether they are 2-component, single component systems, or if the fit failed.}
\label{fig:highbt}
\end{minipage}
\hfill
\begin{minipage}[b]{0.47\textwidth}
\includegraphics[width=\textwidth]{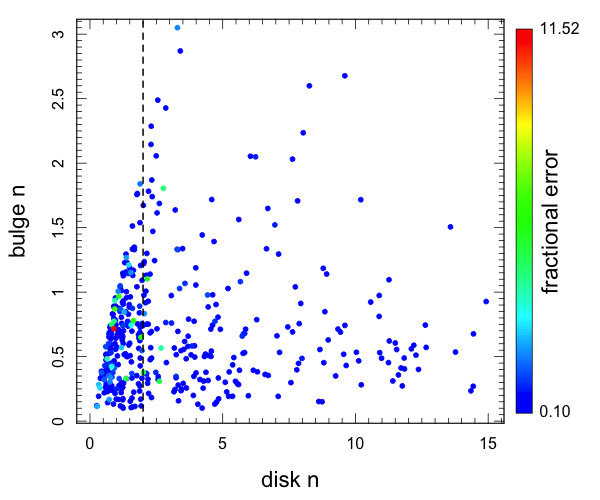}
\caption{The disc versus bulge S\'ersic index distribution for galaxies with bulge n $<$ disc n.
The points are colour coded by the fractional error on the disc S\'ersic index.
It is obvious that many of the fits for these galaxies converged (i.e. 10$\%$ error).
We remove all converged and disc n $>$ 2 galaxies from our final sample since we 
consider their fits unphysical.}
\label{fig:nreverse}
\end{minipage}
\end{figure}

Fig.~\ref{fig:highbt} shows the distribution of high \bt{} galaxies. We set all 46 galaxies with a \bt{} $>$ 0.8 to a single component fit. 
Additionally, since our mean \bt{} error is 0.1, we also consider all galaxies with a \bt{}$<$0.1 to be a single component galaxy. This adds another 42 galaxies for which the 2-component fit is considered not appropriate.\\

\noindent \textbf{(ii) disc S\'ersic index $>$ bulge S\'ersic index}\\
As mentioned in Sec.~\ref{sec:switch} during the set up of our final \bt{} decomposition
catalogue we screen galaxies for bulge and disc component switching.
To do this we assume that the bulge R$_e$ is smaller than the disc R$_e$ and swap
the assigned component for those galaxies where this is not the case.
However, for galaxies where the bulge R$_e$ is up to 10$\%$ larger than the disc
R$_e$ we also check whether the disc n is larger than the bulge n and only
swap the assigned component if this is the case.

However, even though the swapping ensures that our bulge is smaller than the disc
it is not guaranteed that the bulge n is larger than the disc n.
We check our 2-component fits and find 443 galaxies where the bulge n is smaller
than the disc n. Fig.~\ref{fig:nreverse} shows the bulge versus disc n distribution coloured by the
percentage error for the 443 flagged galaxies. 
Visually inspecting a number of the resulting fits we find that
galaxies with a disc $n>2$ and/or converged fits are typically bad and we remove
them from consideration of the component samples. In total we remove 100 S(B)0-S(B)0a
galaxies and 215 S(B)ab-S(B)cd galaxies.\\

\begin{figure}
\centering
\begin{minipage}[b]{0.45\textwidth}
 \includegraphics[width=\textwidth]{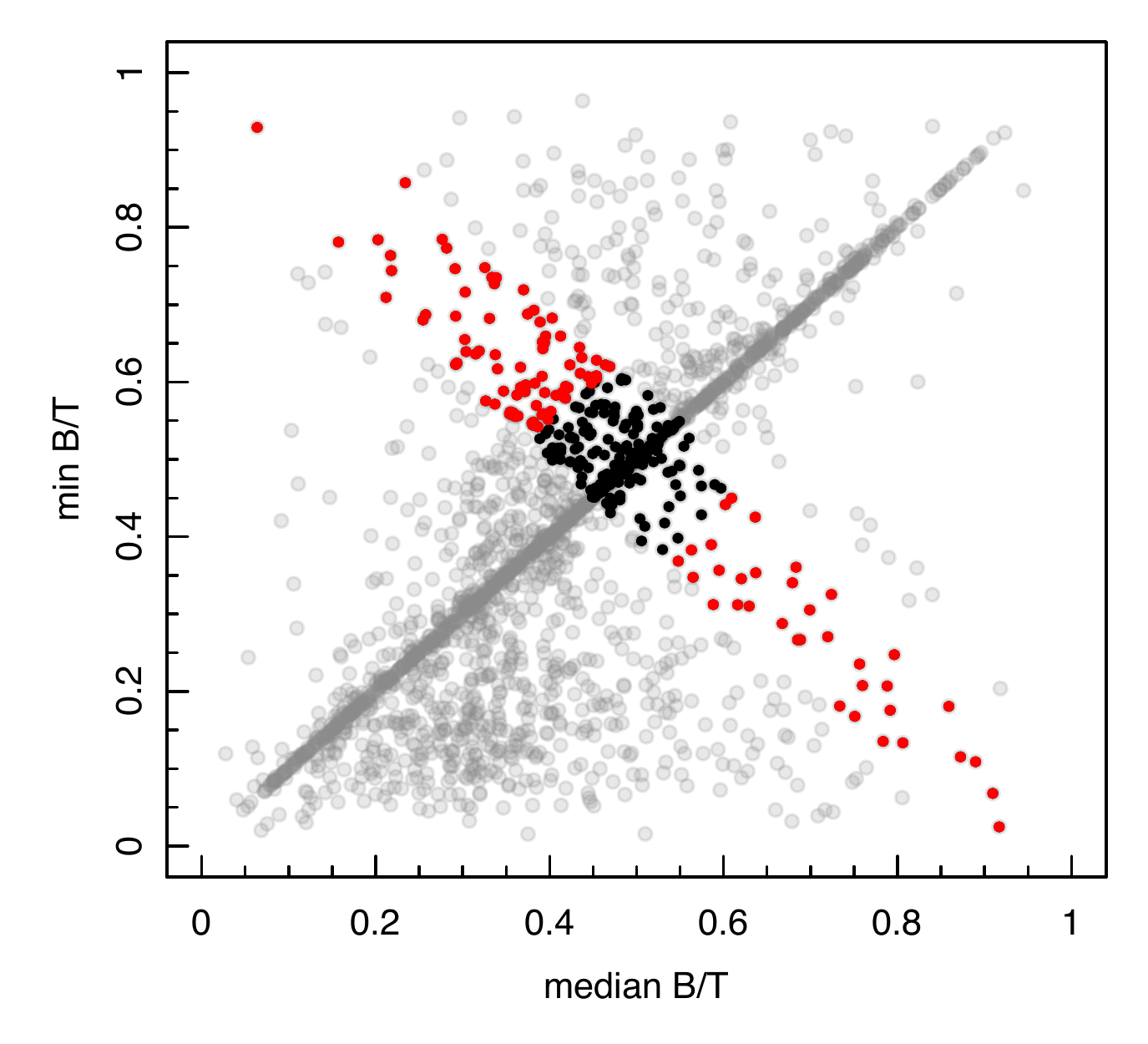}
   \caption{The median vs minimum reduced \x{} distribution of the \bt{} in grey. The black points show
the galaxies for which the sum of the min and median models lies between 0.9 and 1.1. The red points
show the galaxies for which additionally the absolute difference between the min 
and median models \bt{} is larger than 0.15.}
\label{fig:btreverse}
\end{minipage}
\hfill
\begin{minipage}[b]{0.45\textwidth}
 \includegraphics[width=\textwidth]{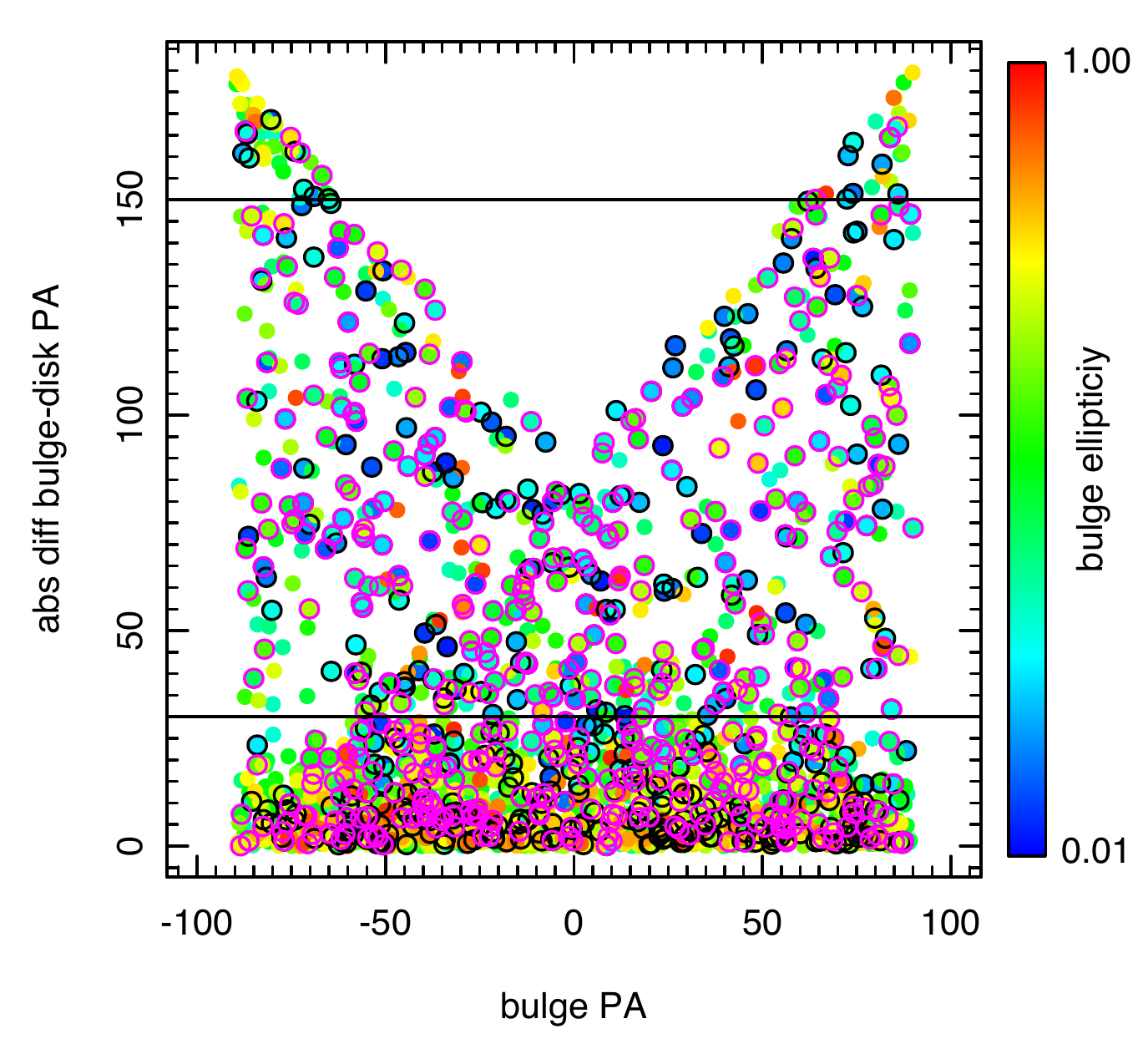}
 \caption{The absolute difference between bulge and disc 
 position angle versus the bulge position angle. The points are coloured 
 by the bulge ellipticity (e=1-b/a). The two vertical lines at 30 and 150 
 degrees show the range between which we flag the galaxies. The black circles 
 show galaxies with bulge e$<$0.3 and the pink circles show galaxies with a disc e$<$0.3, which are excluded from the flagging.}
\label{fig:paflag}
\end{minipage}
  \end{figure}

\noindent \textbf{(iii) bulge and disc fraction reverse between median and minimum reduced \x{} model}\\
We have 116 galaxies for which the median and minimum reduced \x{} models have 
opposite \bt{} values.
To establish this sample we select galaxies where the sum of the median and 
min \bt{} is between 0.9 and 1.1. 
The spread of this sum should include our inherent uncertainty in establishing 
the \bt{}, which is around 0.1-0.15. However, since this criterion alone also flags
 galaxies with a \bt{}=0.5 we add a second criterion that the absolute difference
  between the median and min \bt{}  has to be larger than 0.15 (i.e.~larger than our
   average uncertainty in establishing a \bt{}). Figure \ref{fig:btreverse} shows the \bt{} 
   distribution for the median versus minimum reduced \x{} model, and highlights the 
   process of flagging the galaxies with a reversed \bt{} between the median and
    minimum reduced \x{} models.\\

\noindent \textbf{(iv) bulge and disc position angle more than 30 degrees offset}\\
This flag is defined by identifying all galaxies for which the bulge and disc position angle (PA) differs by more than 30 degrees.
However, for round bulges or discs the PA is not very meaningful
and even a large offset between bulge and disc PA is not indicative of a problem 
with the fitting results.
To exclude these galaxies we set a second condition for this flag, namely that the bulge or disc ellipticity has to be large enough to have a clearly identifiable preferred major axis, i.e. e=1-b/a$>$0.3. 
This reduces the flagged galaxies to 100 objects.\\
 Figure \ref{fig:paflag} shows the distribution of the absolute PA offset vs bulge PA,
  with the points coloured by their ellipticity.
 The horizontal lines show the angle offset between which we flag, and the objects
  circled show galaxies with e$<$0.3 which are excluded from the flagging.\\

\begin{figure*}
\centering
\includegraphics[width=0.9\textwidth]{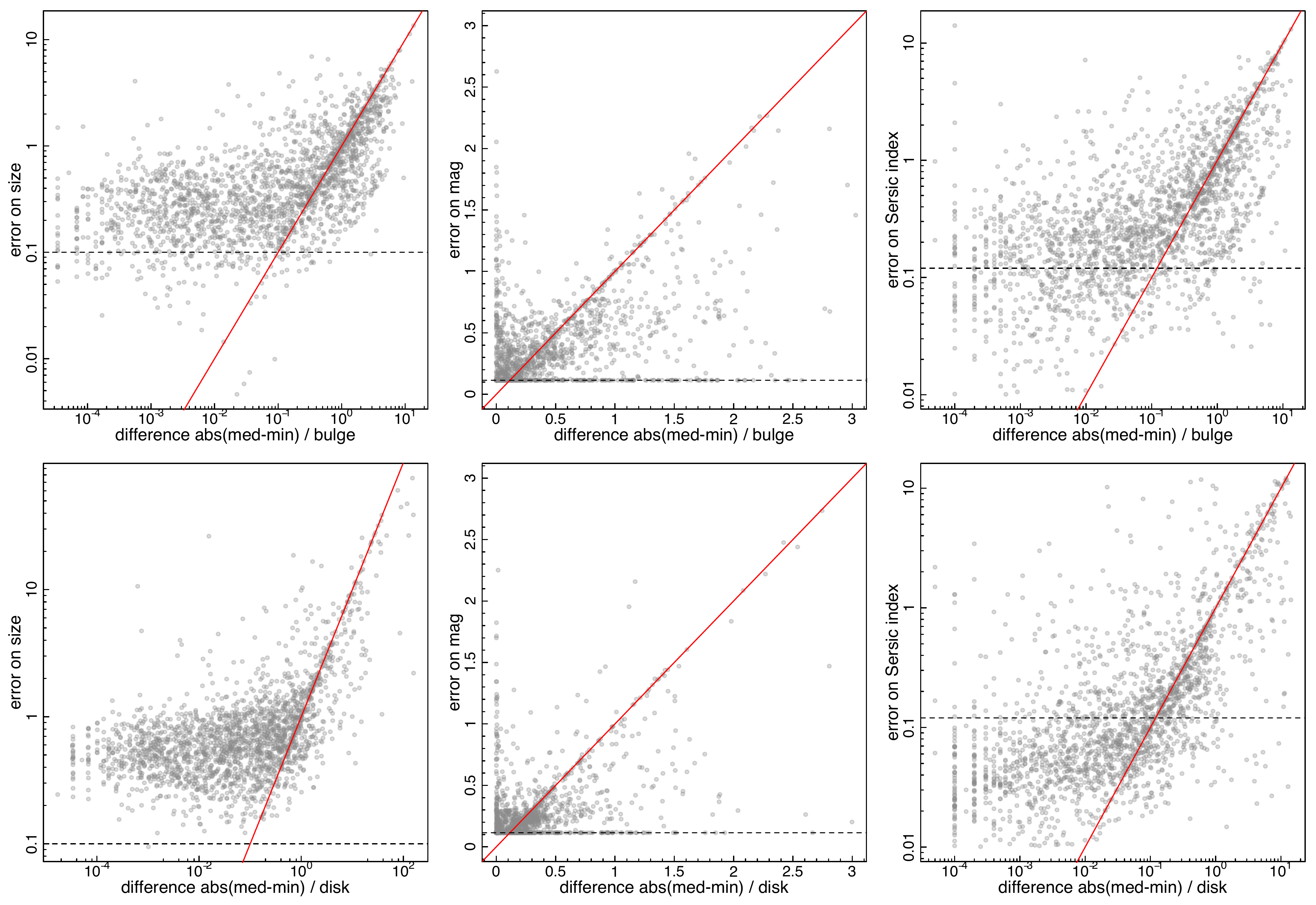}
\caption{The top panels show bulge values, on the bottom panels the disc values.
From left to right we show: \newline
difference in median and minimum reduced \x{} model sizes versus the median size error \newline
difference in median and minimum reduced \x{} model magnitudes versus the median magnitude error \newline
difference in median and minimum reduced \x{} model S\'ersic indices versus the median S\'ersic index error \newline
The red lines show the 1:1 correspondence, the black dashed lines indicate our lower limits on
the errors. Everything to the right of the red lines and above the dashed lines is flagged.}
\label{fig:outlierflag}
\end{figure*}

\noindent \textbf{(v) minimum reduced \x solution represents an outlier to the median model}\\
Finally we flag galaxies for which the minimum reduced \x solution represents and outlier, i.e. where the error on the median
values does not include the minimum reduced \x{} solution. For minimum reduced \x solutions where the parameters are
smaller than the median values we consider the error on the lower end, and vice versa where the median is smaller 
we consider the error on the higher end.\\
We tested size, magnitude and S\'ersic index distributions for outliers. We only consider the minimum reduced \x{} values to be a true
outlier, and possibly better fit, if both the disc and bulge values fall outside the error range. Additionally we check that 
the error on the median is larger than our assumed error floor, otherwise we consider the various fits of the galaxy to have 
converged and thus the minimum reduced \x solution likely represents a failed fit.\\
Figure \ref{fig:outlierflag} shows the distribution of the absolute difference between the median and minimum reduced \x{}
values versus the error for the median values. The left hand panel shows the bulges and the right hand panel the
disc distribution. From top to bottom we investigate size, magnitude, and S\'ersic index.\\
The red lines show the 1:1 correspondence, i.e. where the minimum reduced \x{} value lies on the edge of the median error
distribution. The horizontal dashed lines shows our error floor, and we assume that most models converged to give
rise to such a low error.\\
All galaxies to the right of the red lines and above the dashed black lines are potentially bad. We do, however,
only flag the ones that are bad for both the disc and bulge values. This gives rise to
54 galaxies with flagged sizes, 393 galaxies with flagged magnitudes, and 88 galaxies 
with flagged S\'ersic indices.
In total this results in 449 galaxies for which the minimum reduced \x solution is an outlier 
to at least one of the three fitting parameters tested.\\

The equivalent test on our single component fits finds 73 (42) spheroid (disc) sizes,
27 (30) spheroid (disc) magnitudes, and 62 (48) spheroid (disc) S\'ersic indices
 flagged. 
In total this equates to 90 (74) spheroid (disc) galaxies with at least one of the
 parameters flagged ($\leq$5).

\section{Simulation comparisons}
\label{app:sim_comp}

\begin{figure*}
\centering
\includegraphics[page=1,height=0.93\textheight]{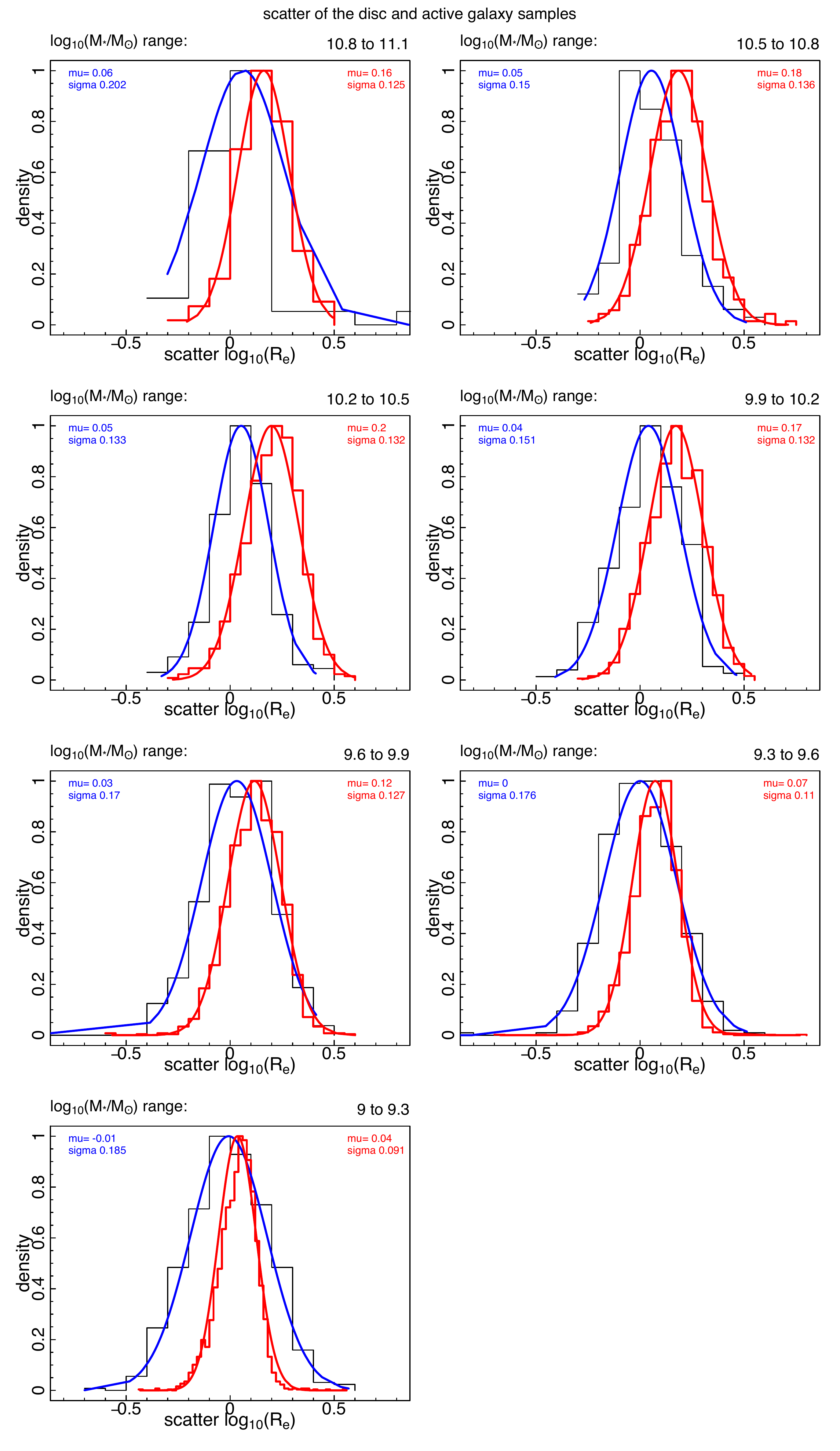}
\captionof{figure}{a) Distribution of the half-light radius scatter of discs for bins in 
stellar mass (shown at the top of each panel) for our data.
The blue line shows the normal distribution fit to the observed data (black histogram) 
and the red line is the normal distribution fit for the
 simulated data (red histogram), with the caveat that we use the half-mass radius for star forming galaxies in the case of \eagle . 
The legend on the left shows the best fit values for the observed data and on the right for the 
simulated data.}
\label{fig:logrescatter}
\end{figure*}

\begin{figure*}
\centering
\includegraphics[page=1,height=0.93\textheight]{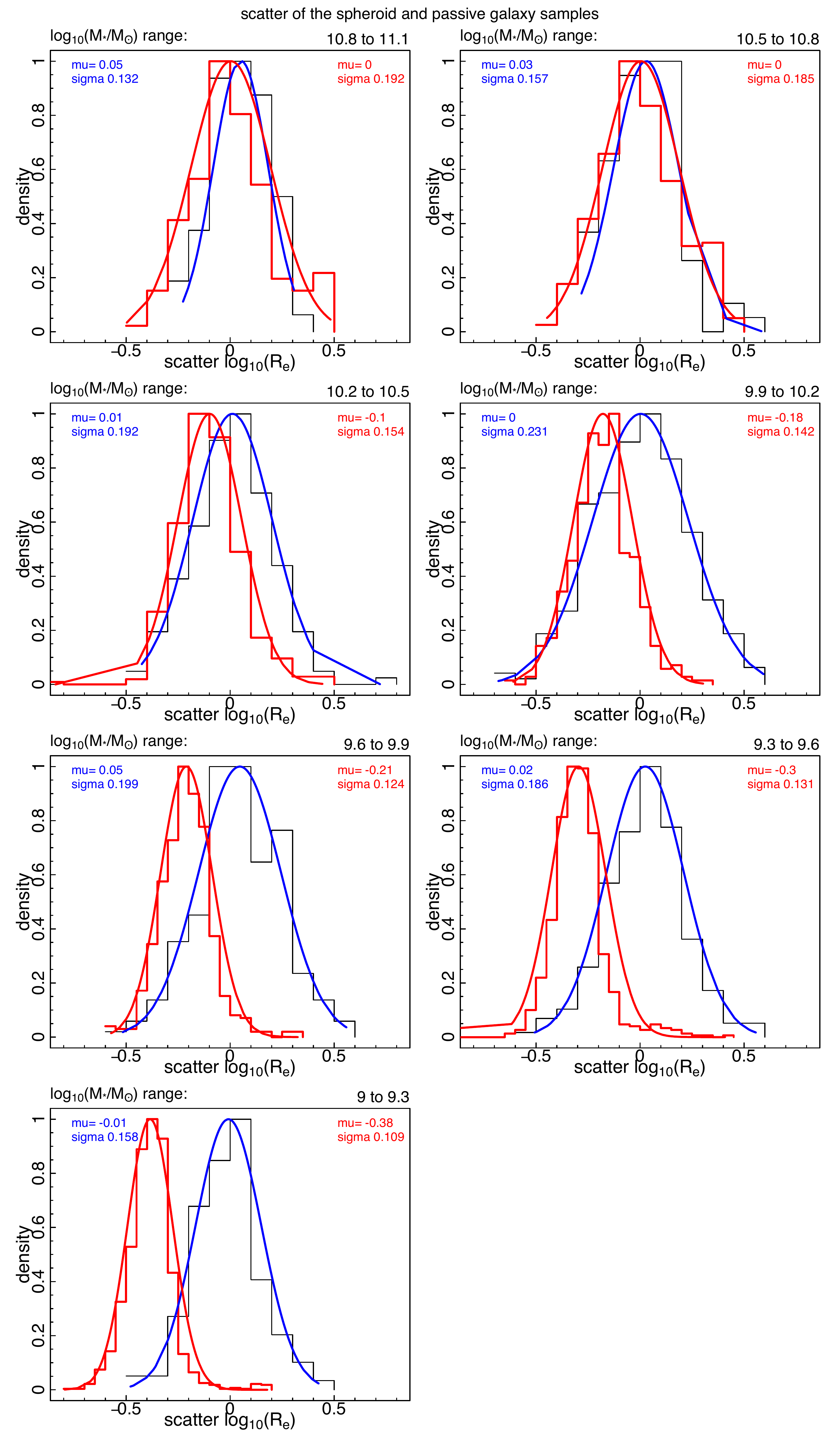}
\caption{ Same as Fig.~\ref{fig:logrescatter} but shown is the $\log(R_e)$ scatter for the spheroid component distribution and in 
the case of \eagle\ we show the scatter of the half-mass radius of passive galaxies.}
\label{fig:logrescatter2}
\end{figure*}

In Figures \ref{fig:logrescatter} and \ref{fig:logrescatter2} we show the distribution of 
the vertical scatter from the \msr{} relation for discs and spheroids. 
The black histogram shows the distribution of our data and the 
blue line is a fit of a normal distribution to it. The red histograms show the distribution 
 of the z=0 \eagle\ data with the red line being the normal distribution 
fit to it. 

Figure \ref{fig:simcomp} is essentially derived from these plots where the standard deviation (SD) points
refer to the sigma of the Gaussian fit and the 16$^{th}$ and 84$^{th}$ percentiles are the width of the
underlying distribution.
Typically these measurements of sigma should 
be the same, however, in some cases the 
distributions have tails towards higher deviations which causes the 16$^{th}$/84$^{th}$ percentile
 and SD measurements to be different (as seen in Fig.~\ref{fig:simcomp}).

\label{lastpage}                                                                                                                                                                                                                                                                                                                                         
\end{document}